\documentclass[twocolumn,useAMS,usenatbib]{mn2e} \usepackage{natbib}

\usepackage{amsmath,latexsym,amssymb} \usepackage{epsfig,graphicx} \topmargin-1cm
\usepackage{epstopdf}
\usepackage{float}
\usepackage{fixltx2e}
\usepackage{color}
\usepackage{url}
\graphicspath{ {./draft3_figs/}}
\DeclareGraphicsExtensions{.pdf,.png,.jpg,.eps}

\def\reff@jnl#1{{\rm#1\/}}

\def\aj{\reff@jnl{AJ}}                  
\def\araa{\reff@jnl{ARA\&A}}            
\def\apj{\reff@jnl{ApJ}}                        
\def\apjl{\reff@jnl{ApJ}}               
\def\apjs{\reff@jnl{ApJS}}              
\def\apss{\reff@jnl{Ap\&SS}}            
\def\aap{\reff@jnl{A\&A}}               
\def\aapr{\reff@jnl{A\&A~Rev.}}         
\def\aaps{\reff@jnl{A\&AS}}             
\def\baas{\reff@jnl{BAAS}}              
\def\jcap{\reff@jnl{JCAP}}              
\def\jrasc{\reff@jnl{JRASC}}            
\def\memras{\reff@jnl{MmRAS}}           
\def\mnras{\reff@jnl{MNRAS}}            
\def\physrep{\reff@jnl{Phys.Rep.}}
\def\pra{\reff@jnl{Phys.Rev.A}}         
\def\prb{\reff@jnl{Phys.Rev.B}}         
\def\prc{\reff@jnl{Phys.Rev.C}}         
\def\prd{\reff@jnl{Phys.Rev.D}}         
\def\prl{\reff@jnl{Phys.Rev.Lett}}      
\def\pasp{\reff@jnl{PASP}}              
\def\pasj{\reff@jnl{PASJ}}              
\def\skytel{\reff@jnl{S\&T}}            
\def\solphys{\reff@jnl{Solar~Phys.}}    
\def\sovast{\reff@jnl{Soviet~Ast.}}     
\def\ssr{\reff@jnl{Space~Sci.Rev.}}     
\def\nat{\reff@jnl{Nature}}             

\newcommand{\hmpc}{\ensuremath{h^{-1}\mathrm{Mpc}}}

\newcommand{\hMsun}{\ensuremath{h^{-1}M_{\odot}}}

\newcommand{\beq}{\begin{equation}}
\newcommand{\eeq}{\end{equation}}
\newcommand{\beqa}{\begin{eqnarray}}
\newcommand{\eeqa}{\end{eqnarray}}

\title[Galaxy Shapes and Alignments]{Galaxy shapes and alignments in the MassiveBlack-II hydrodynamic and dark matter-only simulations}
\author[Tenneti et al.]
{Ananth Tenneti$^1$\thanks{\tt vat@andrew.cmu.edu},
Rachel Mandelbaum$^1$\thanks{\tt rmandelb@andrew.cmu.edu},
Tiziana Di Matteo$^1$\thanks{\tt tiziana@phys.cmu.edu},
\newauthor Alina Kiessling$^2$,
Nishikanta Khandai$^{3}$ 
\\$^1$McWilliams Center for Cosmology, Department of Physics, Carnegie Mellon University, Pittsburgh, PA 15213, USA
\\$^2$Jet Propulsion Laboratory, California Institute of Technology, Pasadena, CA 91109, USA
\\$^3$School of Physical Sciences, National Institute of Science Education and Research, Bhubaneswar, Odisha 751005, India} 

\date{\today}
\begin{document}
\maketitle

\begin{abstract}
  We compare the shapes and intrinsic alignments of galaxies 
  in the MassiveBlack-II cosmological hydrodynamic simulation (MBII) to
  those in a dark matter-only (DMO) simulation performed
  with the same volume (100$h^{-1}$Mpc)$^{3}$, cosmological parameters, and initial
  conditions. Understanding the impact of baryonic physics on galaxy shapes and alignments and their relation to the
  dark matter distribution should prove useful to map
  the intrinsic alignments of galaxies from hydrodynamic to dark matter-only simulations. We find
  that dark matter subhalos are typically rounder in MBII, and the shapes of
  stellar matter in low mass galaxies are more misaligned with the
  shapes of the dark matter of the corresponding subhalos in the DMO
  simulation. At $z=0.06$, the fractional difference in the mean misalignment angle between MBII and DMO simulations
  varies from $\sim 28 \% - 12 \%$ in the mass range
  $10^{10.8} - 6.0 \times 10^{14} h^{-1}M_{\odot}$. We study the dark matter halo shapes and
  alignments as a function of radius, and find that while galaxies in MBII are more aligned with the
  inner parts of their dark matter subhalos, there is no radial trend in their alignments with the
  corresponding subhalo in the DMO simulation. This result highlights the importance of baryonic
  physics in determining the alignment of the galaxy with respect to the inner parts of the halo.
Finally, we compare the ellipticity-direction (ED) correlation for galaxies
to that for dark matter halos, finding that it is
suppressed on all scales by stellar-dark matter misalignment. In the
projected shape-density
correlation ($w_{\delta+}$), which includes ellipticity weighting, this effect is partially
canceled by the higher mean ellipticities of the stellar component, but
differences of order $30-40\%$ remain on scales $> 1$ Mpc over a range of subhalo masses, with
scale-dependent effects below $1$ Mpc.
\end{abstract}


\section{Introduction} \label{S:intro}

 Weak gravitational lensing is a cosmological probe that has the
 potential to address major outstanding cosmological problems, such as understanding
the connection between dark matter and galaxies, the nature of dark
energy, and exploring possible variations in the theory of gravity on
cosmological scales \citep{{2006astro.ph..9591A},2013PhR...530...87W}. Using weak
lensing, upcoming surveys such as the Large Synoptic Survey Telescope\footnote{\url{http://www.lsst.org/lsst/}}
(LSST; \citealt{LSST09}),
Euclid\footnote{\url{http://sci.esa.int/euclid/},
  \url{http://www.euclid-ec.org}} \citep{LAA+11}, and the Wide-Field 
Infrared Survey Telescope\footnote{\url{http://wfirst.gsfc.nasa.gov}}
(WFIRST; \citealt{SGB+15}) will to constrain
cosmological parameters such as the dark energy equation of state
to a very high precision. However, the most basic weak lensing analysis assumes that
 galaxy shapes are randomly aligned, which is not correct in
reality. The galaxy shapes are correlated with each other and with the
underlying density field. This intrinsic alignment of galaxy shapes 
is an important astrophysical systematic 
that contaminates measurements in weak lensing surveys
\citep{{2000MNRAS.319..649H},{2000ApJ...545..561C},{2001MNRAS.320L...7C},{2002MNRAS.335L..89J},{2004PhRvD..70f3526H}}. For reviews on intrinsic alignments and its impact on 
cosmological parameter constraints, see \cite{2014arXiv1407.6990T},
\cite{2015arXiv150405456J}, \cite{2015arXiv150405465K}, and \cite{2015arXiv150405546K}.
 
Previous studies of intrinsic alignments that made predictions
  out to tens of Mpc scales have generally involved analytical methods, or $N$-body and hydrodynamic
simulations of cosmological volume. Analytical modeling
involves a number of different possible models, with the one receiving
the most attention for elliptical galaxies being the linear alignment model \citep{{2001MNRAS.320L...7C},{2004PhRvD..70f3526H}}, an
extension of it to nonlinear scales using the non-linear power
spectrum (\citealt{2007NJPh....9..444B}; see \citealt{2015arXiv150402510B} for a more
recent extension that includes all non-linear terms at the same
order), and a small-scale extension using the halo model 
\citep{2010MNRAS.402.2127S}. $N$-body simulations have been used to
study intrinsic alignments by
\cite{2006MNRAS.371..750H} where halos are populated with galaxies which are stochastically misaligned, \cite{2013MNRAS.436..819J} with semi-analytic models, and others
(see references in the review by \citealt{2015arXiv150405546K}). However, the physics
of galaxy formation is not taken into consideration by these analytic
models and $N$-body simulations. Recent hydrodynamic simulations of
cosmological volumes such as the MassiveBlack-II
\citep{2014arXiv1402.0888K}, the Horizon-AGN
\citep{2014MNRAS.444.1453D}, and the EAGLE and cosmo-OWLS
\citep{2015arXiv150404025V} simulations have made it possible to
directly study 
the intrinsic alignments of the stellar component of galaxies and their
scaling with mass, luminosity and distance. 

Using the MassiveBlack-II (MBII) simulation, we previously studied the
shapes of the stellar matter component of the galaxies and its
alignment with the shape of the dark matter component
\citep{2014MNRAS.441..470T}. We found that the ellipticities of the
stellar components of galaxies compare favorably with those in observations, and
that the shape of the stellar component is misaligned with the dark matter
component, with average misalignment angles ranging from $\sim
10^{\circ}$ to $30^{\circ}$. In
\cite{2015MNRAS.448.3522T}, we extended this study to two-point
statistics and explored their dependence with mass, luminosity and
distance. We found that the intrinsic alignments of massive galaxies
in the simulation have a scaling with transverse distance that is
consistent with results from observational measurements. Large volume 
hydrodynamic simulations are proving to be extremely useful in 
quantifying intrinsic alignment signals. However,
they are also very computationally 
expensive. Hence, it would be useful to develop methods to paint the
intrinsic alignments of galaxies onto $N$-body simulations. It is
known that the halo shapes and orientations in $N$-body simulations overestimate the intrinsic
alignments when compared with observational results
\citep{2006MNRAS.371..750H}. Hence, a direct comparison of intrinsic
alignments in hydrodynamic and $N$-body simulations is necessary. In
this paper, we use the MassiveBlack-II dark matter-only (DMO)
simulation that has been performed with the same resolution, box size,
initial conditions,
and cosmological parameters as the hydrodynamic simulation (MBII)
to compare the intrinsic alignments of galaxies in MBII and halos in DMO.

To predict the intrinsic alignment of galaxies from dark matter-only
 simulations accurately, it is important to have an average mapping
 that statistically determines the ellipticity
and orientation of the stellar matter component of the subhalo with
respect to that of the dark matter component. Here, the radius at which the shape
of the dark matter subhalo is measured matters, as it has
been shown using $N$-body simulations that the shapes of dark matter
component change with radius
\citep{2006MNRAS.367.1781A,2007ApJ...671.1135K,2012JCAP...05..030S}. 
Further, it is also known that the dark matter halo shapes in hydrodynamic
simulations are different from those in $N$-body simulations due to the
effects of baryonic physics, which leads to rounder shapes
\citep{2004ApJ...611L..73K,2013MNRAS.429.3316B}. Here, we study the
radial dependence of the distribution of dark matter halo axis ratios
in the MBII and DMO
simulations, and compare with the shape of the stellar
matter component in MBII. We also study the orientation of the stellar
shape with the shape of dark matter component in MBII and DMO measured
at different radii. The change in
the orientation of the dark matter shapes in the subhalos measured at
different radii has been previously studied using $N$-body simulations \citep[e.g.,][]{{2005ApJ...627..647B,2012JCAP...05..030S}}. Using small volume
hydrodynamic simulations, \cite{2005ApJ...627L..17B},
\cite{2011MNRAS.415.2607D} and others (see references in the review by \citealt{2015arXiv150405546K}) found that the inner shape of the dark
matter component is well aligned with the galaxy orientation. However, it will be useful to compare the orientation
of stellar shape in the hydrodynamic simulation with the shape of the
dark matter component measured at different radii in the matched
subhalo of a dark matter-only simulation.  This comparison is what
will enable the mapping of intrinsic alignments in hydrodynamic
simulations onto dark matter-only simulations.

This paper is organized as follows. In Section~\ref{S:methods}, we
describe the details of the simulations used in this study followed by
the method adopted to determine the shapes of galaxies and the
definitions of two-point statistics. In Section~\ref{S:results}, we
match the subhalos in both simulations and compare shapes and
misalignment angles. A brief discussion of the comparison of mass
functions and dark matter power spectrum in the two simulations is
also included here. In Section~\ref{S:radialshapes}, we analyze the
radial dependence of shapes and orientations of the dark matter
component in subhalos. In Section~\ref{S:edwdp}, we compare the
two-point correlation functions using shapes defined by the dark
matter component in MBII and DMO, and the stellar shapes from
MBII. Finally, we summarize our conclusions in
Section~\ref{S:conclusions}.
             
\section{Methods}\label{S:methods}

\begin{figure*}
\begin{center}
\includegraphics[width=3.2in]{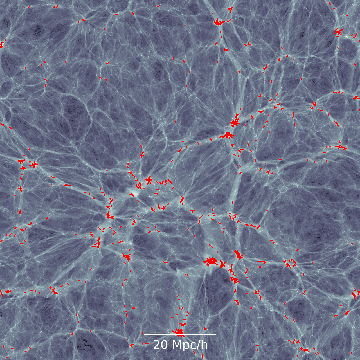}
\includegraphics[width=3.2in]{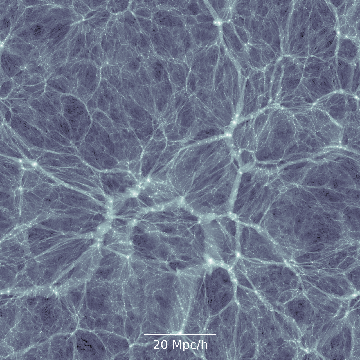}
\caption{\label{F:fig_lssDM} Snapshot of the MBII (left) and DMO (right) simulations in a slice of thickness
  $2h^{-1}$Mpc at $z=0.06$. The shading represents the density distribution of dark matter, and
  the red lines show the alignment of stellar shapes in MBII. The length of the lines is proportional to the size of major axis of the ellipse representing stellar shape.}
\end{center}
\end{figure*}

\begin{figure*}
\begin{center}
\includegraphics[width=3.2in]{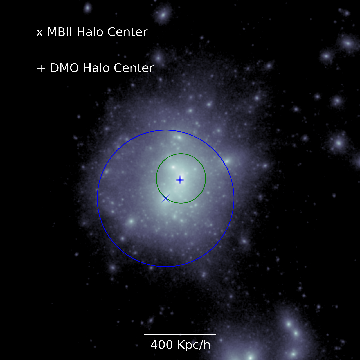}
\includegraphics[width=3.2in]{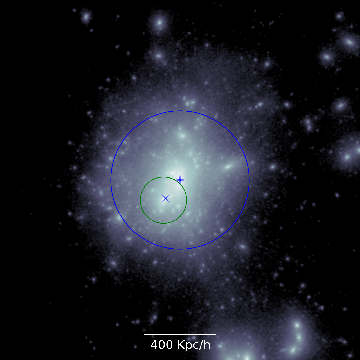}
\caption{\label{F:fig_halo0DM} Snapshot of a massive halo ($\sim 10^{14} \hMsun$) in the MBII (left)
  and DMO (right) simulations, showing the
  density distribution of dark matter at $z=0.06$. The blue and green circles show the virial radii
  of the subhalos centered at the location with highest density. In both the panels, 'x' and '+'
  indicate the locations of the halo centers in the MBII and DMO simulations, respectively.}
\end{center}
\end{figure*}

\subsection{Simulations}\label{SS:sims}
\begin{table}
\caption{\label{T:param}Simulation parameters: box size (L$_\text{box}$), force softening length ($\epsilon$), number of particles ($N_\text{part}$), mass of dark matter particle ($m_\text{DM}$) and mass of gas particle ($m_\text{gas}$)}
\begin{tabular}{|c|c|c|}
\hline
 Parameters & Hydrodynamic & Dark Matter Only \\ & (MBII) & (DMO)\\
\hline  
L$_\text{box}$ ($h^{-1}$Mpc)& 100 & 100\\ 
$\epsilon$ ($h^{-1}$kpc)& 1.85 & 1.85\\
$N_\text{part}$ & $2 \times 1792^{3}$ & $1792^{3}$\\
$m_\text{DM}$ ($h^{-1}M_{\odot}$) & $1.1 \times 10^{7}$ & $1.32 \times 10^{7}$\\
$m_\text{gas}$($h^{-1}M_{\odot}$) & $2.2 \times 10^{6}$ & $0$\\
\hline
\end{tabular}
\end{table}

In this study, we use the MassiveBlack-II hydrodynamic (MBII) and Dark
Matter Only (DMO) \cite{2014arXiv1402.0888K} simulations. 
MassiveBlack-II is a state-of-the-art high resolution, large volume,
cosmological simulation of structure formation. These simulations have
been performed in a cubic periodic box of size $100$\hmpc\ on a side
using {\sc p-gadget}, which is a hybrid version of the parallel code,
{\sc gadget2} \citep{2005MNRAS.361..776S} that has been upgraded to
run on Petaflop-scale supercomputers. The total number of dark matter
particles in both the simulations is $1792^{3}$ with an equal
initial number of gas particles in the hydrodynamic run. The cosmological parameters are chosen according to WMAP7
\citep{2011ApJS..192...18K}, with amplitude of matter
fluctuations $\sigma_{8} = 0.816$, spectral index $n_{s} = 0.96$,
matter density parameter $\Omega_{m} = 0.275$, cosmological constant
density parameter $\Omega_{\Lambda} = 0.725$, baryon density parameter
$\Omega_{b} = 0.046$ (in MBII), and Hubble parameter $h = 0.702$. 
Table~\ref{T:param} shows the box size (L$_\text{box}$), force
softening length ($\epsilon$), total number of particles including
dark matter and gas ($N_\text{part}$), mass of dark matter particle
($m_\text{DM}$) and initial mass of gas particle ($m_\text{gas}$) for
the two simulations.  MBII includes the physics of a multiphase interstellar 
medium (ISM) model with star formation \citep{2003MNRAS.339..289S}, black hole
accretion and feedback \citep{2005MNRAS.361..776S,2012ApJ...745L..29D}
in addition to gravity and smoothed-particle hydrodynamics
(SPH). Radiative cooling and heating processes are included \citep[as
in][]{1996ApJS..105...19K}, as is photoheating due to an imposed
ionizing UV background. Further details about the MBII simulation can
be found in \cite{2014arXiv1402.0888K}. The dark matter-only
simulation, DMO, is performed with the same volume, resolution,
cosmological parameters and initial conditions as in
MBII. Figure~\ref{F:fig_lssDM} shows snapshots of the dark matter
distribution in a slice of $2h^{-1}$Mpc thickness at $z=0.06$ in both the MBII and DMO
simulations. As expected, the dark
matter distributions appear to be nearly identical on large scales. For further
comparison, we also show the dark matter distribution in an isolated
halo in Figure~\ref{F:fig_halo0DM}. Here, we can clearly see small 
differences in the distribution of dark matter within the dark matter
halo virial radius which, as we shall see,
can lead to changes in the shapes and orientations of the halos and
subhalos.

The halo catalogs in the simulations are generated using the friends
of friends (FoF) halo finder algorithm \citep{1985ApJ...292..371D}
with a linking length of $0.2$ times the mean interparticle
separation. The subhalo catalogs are generated using the {\sc subfind}
code \citep{2001MNRAS.328..726S} on the halo catalogs. Here subhalos
are locally overdense, self-bound groups of particles within the
halo. Groups of particles
are identified as subhalos if they have at least $20$ gravitationally
bound particles. When performing our analysis, however, we use only subhalos from the
hydrodynamic simulation with at least $1000$ dark matter and star particles based on the convergence test in \cite{2014MNRAS.441..470T}.

\subsection{Shapes of dark matter subhalos}\label{SS:shapedef}
In this section, we describe the measurements of dark matter and
stellar matter component shapes in subhalos. We model these shapes as
ellipsoids in three dimensions using 
the eigenvalues and eigenvectors of the reduced inertia tensor. The
projected shapes are calculated by projecting the halos and subhalos
onto the $XY$ plane and modeling the shapes as ellipses. In 3D, 
the eigenvectors of the inertia tensor are
${\hat{e}_{a},\hat{e}_{b},\hat{e}_{c}}$ with corresponding
eigenvalues ${\lambda_{a},\lambda_{b},\lambda_{c}}$, where
$\lambda_{a} > \lambda_{b} > \lambda_{c}$. The eigenvectors represent
the principal axes of the ellipsoid, with the lengths of the principal
axes $(a,b,c)$ given by the square roots of the eigenvalues
$(\sqrt{\lambda_{a}},\sqrt{\lambda_{b}},\sqrt{\lambda_{c}})$. The 3D
axis ratios are defined as
\begin{equation} \label{eq:axisratios}
q = \frac{b}{a}, \,\, s = \frac{c}{a}
\end{equation}

In 2D, the eigenvectors are ${\hat{e}_{a}',\hat{e}_{b}'}$ with 
corresponding eigenvalues ${\lambda_{a}',\lambda_{b}'}$, where
${\lambda_{a}' > \lambda_{b}'}$. The lengths of the major and minor axes
are $a' = \sqrt{\lambda_{a}'}$ and $b' = \sqrt{\lambda_{b}'}$ with axis
ratio $q_{2d} = b'/a'$.

The shapes are determined using an iterative method based on the
reduced inertia tensor:
\begin{equation} \label{eq:redinertensor}
\widetilde{I}_{ij} = \frac{\sum_{n} m_{n}\frac{x_{ni}x_{nj}}{r_{n}^{2}}}{\sum_{n} m_{n}}
\end{equation}
where
\begin{equation} \label{eq:rn2}
 r_{n}^{2} = \sum_{i}x_{ni}^{2}
\end{equation}
This definition of inertia tensor gives more weight to particles that are
closer to the center which is desirable since it eliminates the bias due to loosely bound particles present in the outer regions of the subhalo. In the iterative method, we first determine the axis ratios by the
standard definition of the reduced inertia tensor using all the
particles of a given type in the subhalo. Keeping the enclosed volume
constant \citep[as in][]{2012JCAP...05..030S}, we rescale the lengths
of the principal axes of ellipsoids accordingly.  After this
rescaling, we determine the shapes again, discarding particles outside
the ellipsoidal volume. This process is repeated until convergence is
reached. Our convergence criterion is that the fractional change in
axis ratios must be below 1\%. We considered the impact of using
different definitions of the inertia tensor on the distributions of
shapes and the intrinsic alignment two-point correlation functions in
a previous study \citep{2015MNRAS.448.3522T}.

\subsection{Misalignment angle}\label{SS:ma}
To study the relative orientation between the shapes defined by the 
dark matter and stellar matter components in subhalos, we compute 
the probability distributions of misalignment angles as in \cite{2014MNRAS.441..470T}. 
If $\hat{e}_{da}$ and $\hat{e}_{ga}$ are the major axes of the shapes
defined by the dark matter and stellar matter components,
respectively, then we 
can define the misalignment angle by
\begin{equation} \label{eq:misalignangle}
 \theta_{m} = \arccos(\left|\hat{e}_{da} \cdot \hat{e}_{ga}\right|).
\end{equation}

In previous work \citep{2014MNRAS.441..470T,2015MNRAS.448.3522T}, we
studied the distribution of misalignment angles between the shapes of
the stellar component and dark matter component in the subhalos of
MBII. Here, our aim is to study the misalignment angles between the
shapes of stellar component in MBII and the dark matter component in
the corresponding subhalo of the DMO simulation. Since the simulations
have been performed with the same initial conditions, this is a
well-defined comparison, and the resulting distributions will be
helpful to paint galaxies onto $N$-body simulations by providing the
necessary probability distributions from which to draw the galaxy
orientation.

\subsection{Two-point statistics}\label{SS:pes}
To quantify the intrinsic alignments of galaxies with the large-scale
density field, we use the ellipticity-direction (ED) and the
projected shape-density ($w_{\delta +}$) correlation functions. 
The ED correlation quantifies the position angle alignments of galaxies
in 3D, while the projected shape correlation function can be used to compare
against observational measurements that include the 2D shape of the galaxy.

The ED correlation function cross-correlates the orientation of the
major axes of subhalos with the large-scale density
field. For a subhalo centered at position \textbf{x} with major axis
direction $\hat{e}_{a}$, let the unit vector in the direction of a
tracer of the 
matter density field at a distance $r$ be $\hat{\textbf{r}} =
\textbf{r}/r$. Following the notation of \cite{2008MNRAS.389.1266L},
the ED cross-correlation function is given by
\begin{equation} \label{eq:ED3d}
 \omega(r) = \langle \mid \hat{e}_{a}(\textbf{x})\cdot \hat{\textbf{r}}(\textbf{x}) \mid^{2} \rangle - \frac{1}{3}
\end{equation}
which is zero for galaxies randomly oriented according to a uniform distribution.

The matter density field can be represented using either the positions
of dark matter particles (in which case the correlation function is
denoted by the symbol $\omega_{\delta}$) or the positions of subhalos
(in which case it includes a factor of the subhalo bias, and is simply
denoted $\omega$). Here, we only use $\omega_{\delta}$ to eliminate
the effect of subhalo bias.

The projected shape correlation functions are computed to directly
compare our results from simulations with observations. Here, we
follow the notation of \cite{2006MNRAS.367..611M} to define 
the galaxy-intrinsic shear correlation function,
$\hat{\xi}_{g+}(r_{p},\Pi)$ and the corresponding projected
two-point statistic, $w_{\delta +}$. Here, $r_{p}$ is the comoving
transverse separation of a pair of galaxies in the $XY$ plane and
$\Pi$ is their separation along the $Z$ direction.

If $q_{2d} = b'/a'$ is the axis ratio of the projected shape of
the dark matter or stellar matter component of a subhalo, and $\phi$ is
the major axis position angle with respect to the reference direction
(tracer of the density field), the components of
the ellipticity are given by
\begin{equation} \label{eq:ellipticity}
 (e_{+},e_{\times}) = \frac{1 - q_{2d}^{2}}{1 +
    q_{2d}^{2}}\left[\cos{(2\phi)},\sin{(2\phi)}\right],
\end{equation}
where $e_{+}$ refers to the radial component and
$e_{\times}$ is the component at $45^{\circ}$.  The
matter-intrinsic shear correlation function cross-correlates the
ellipticity of galaxies with the matter density field,
\begin{equation} \label{eq:gicorr}
 \hat{\xi}_{\delta +}(r_{p},\Pi) = \frac{S_{+}D - S_{+}R}{RR}
\end{equation}
where $S_{+}$ represents the ``shape sample'' which is selected on the basis of a threshold or binning in subhalo mass. The dark matter particles used to trace the density
field form a ``density sample'' denoted by $D$. $S_{+}D$ is the following sum over all
shape sample vs.\ dark matter particle pairs with separations $r_{p}$ and $\Pi$:
\begin{equation} \label{eq:SpD}
 S_{+}D = \sum_{i\neq j\mid r_{p},\Pi}\frac{e_{+}(j\mid
   i)}{2\mathcal{R}},
\end{equation}
where $e_{+}(j | i)$ is the $+$ component of the ellipticity of a
galaxy ($j$) from the shear sample relative to the direction of a tracer of
density field ($i$) selected from the density sample. Here, $\mathcal{R}
= (1 - e_\text{rms}^{2})$ is the shear responsivity that converts from distortion
to shear \citep{BJ02}, with $e_\text{rms}$ being the RMS ellipticity per component of the shape sample. $S_{+}R$ is defined by a
similar equation for the correlation of the data sample with a random
density field distribution to remove observational systematics in the
shear estimates, so we neglect this term here. The $RR$ term in Eq.~\ref{eq:gicorr} is given by the
expected number of randomly-distributed points in a particular $(r_p, \Pi)$ bin around galaxies in the shape sample. 

The projected shape correlation function, 
$w_{\delta +}(r_{p})$ is now
given by
\begin{equation} \label{eq:wgp}
 w_{\delta +}(r_{p}) =
 \int_{-\Pi_\text{max}}^{+\Pi_\text{max}}\hat{\xi}_{\delta +}(r_{p},\Pi)\,\mathrm{d}\Pi
\end{equation}
We calculated the correlation functions over the whole length
of the box ($100\hmpc$) with $\Pi_\text{max} = 50\hmpc$. The projected
correlation functions are obtained via direct summation.

\section{Results}\label{S:results}
\subsection{Mass function}
\begin{figure}
\begin{center}
\includegraphics[width=3.2in]{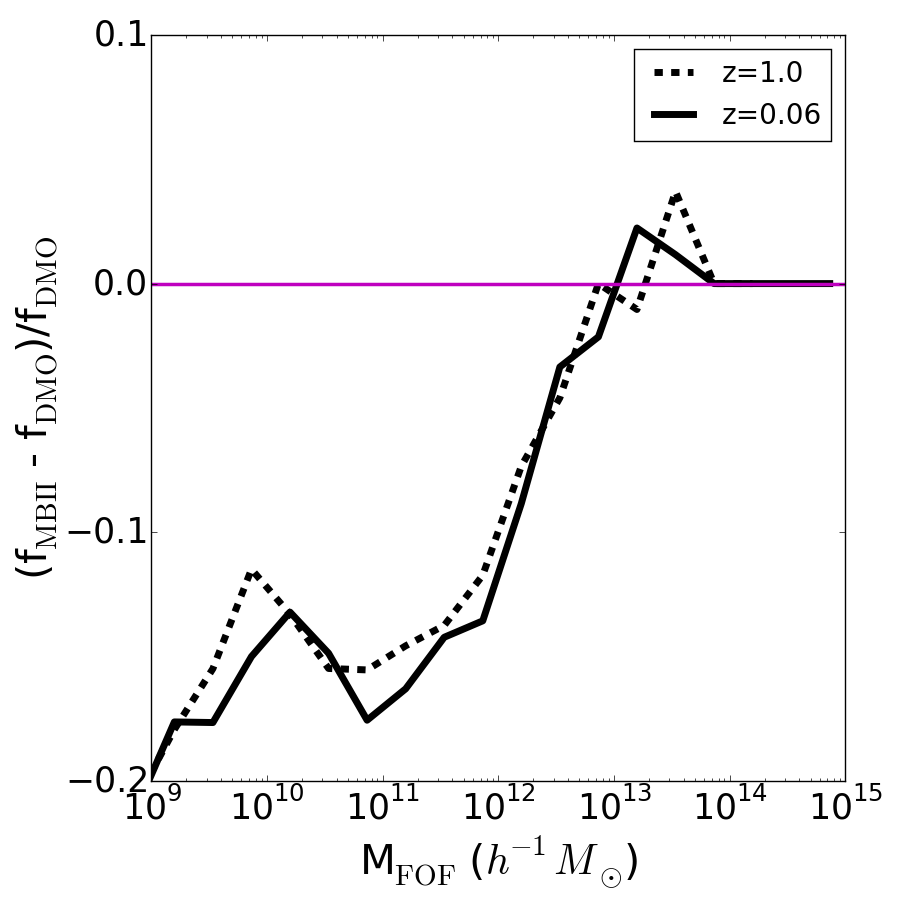}
\caption{\label{F:fig_massf} Comparison of the halo mass function in
  MBII and DMO simulations at $z=1.0$ and $z=0.06$.}
\end{center}
\end{figure}

In Figure~\ref{F:fig_massf}, we compare the mass function of halos in
MBII and DMO simulations at redshifts $z=1$ and $z=0.06$.  At both redshifts, there are differences in the dark matter  halo mass function in
MBII. There are $\sim 10-20 \%$ more halos in the DMO simulation at low masses
($10^{9} - 10^{12} \hMsun$). This fraction decreases at higher
masses. \cite{2014MNRAS.442.2641V} found similar results using the
OWLS simulation with resolved halos in the mass
range $10^{12} - 10^{15} \hMsun$ (based on $M_{200}^\text{mean}$). However, \cite{2014MNRAS.442.2641V} find an even larger
suppression in their halo  mass function, possibly due to a stronger AGN
feedback model. 
\begin{figure}
\begin{center}
\includegraphics[width=3.2in]{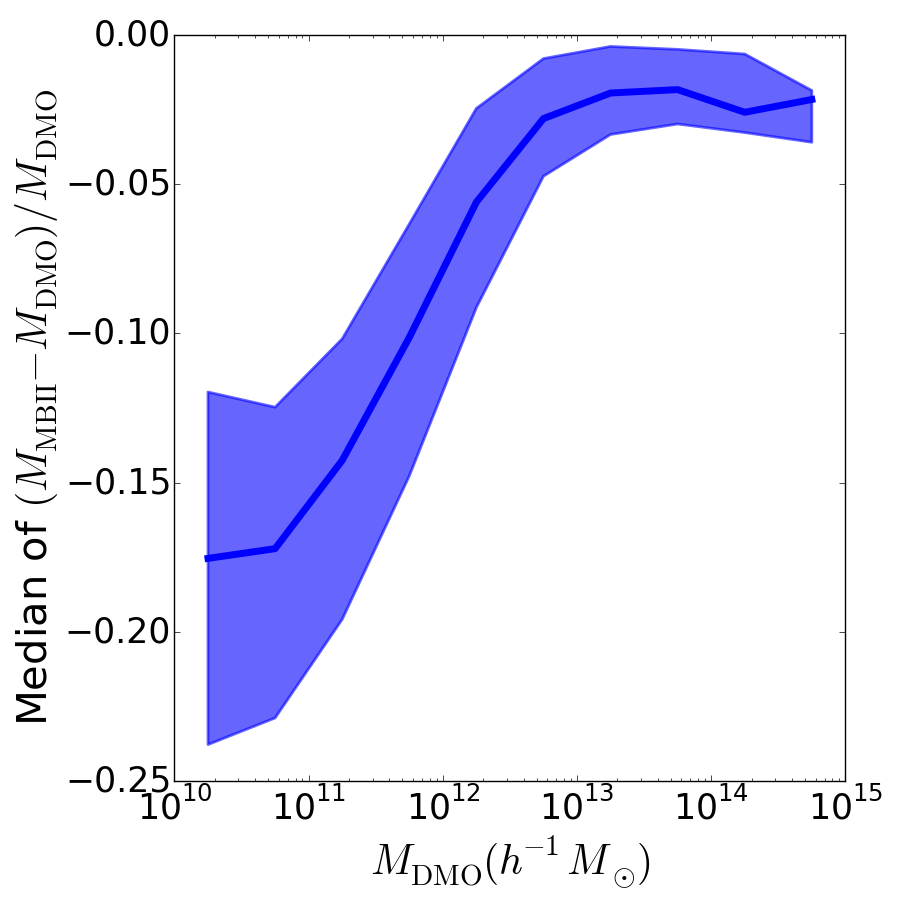}
\caption{\label{F:fig_massdiff} Median and scatter (defined using the $16^\text{th}$ and
  $84^\text{th}$ percentiles) of the fractional difference in the FOF mass of matched halos of MBII and DMO simulations at $z=0.06$.}
\end{center}
\end{figure}
This suppression of mass function can be explained by the decrease in the FOF mass of halos in MBII. To illustrate further, we plotted the median of the fractional difference in the FOF mass of the matched halos of MBII and DMO simulation in Figure~\ref{F:fig_massdiff}. We find that the halo mass is smaller in the hydrodynamic simulation by $\sim 10-20 \%$ in the low mass range.   

\subsection{Matter correlation function and power spectrum}

\begin{figure*}
\begin{center}
\includegraphics[width=3.2in]{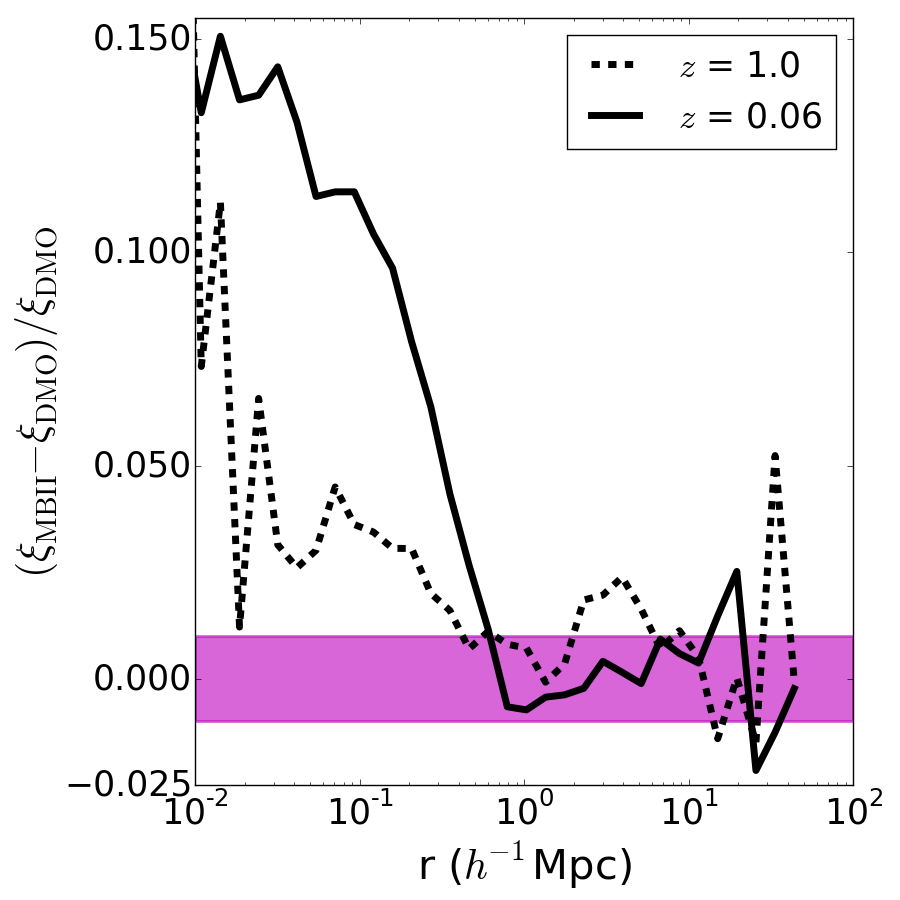}
\includegraphics[width=3.2in]{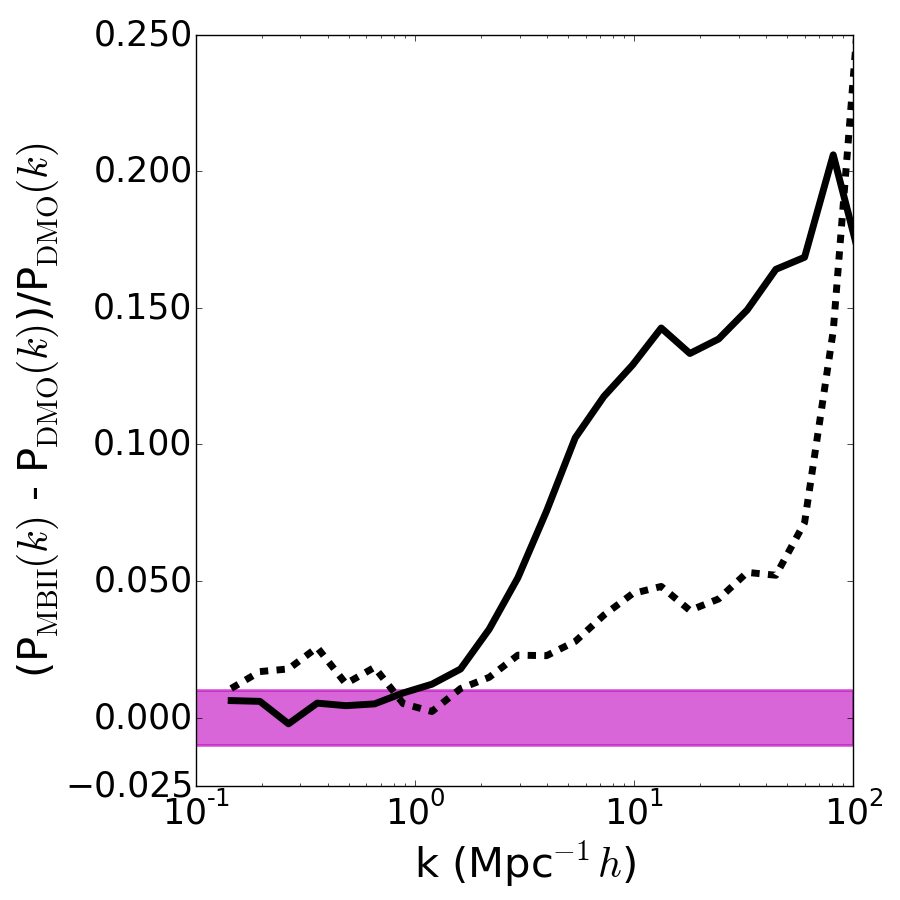}
\caption{\label{F:fig_2pointf} Fractional difference between the dark matter two-point
  correlation functions (left) and power
  spectra (right) in DMO and MBII simulations at $z=1.0$ and $z=0.06$. The shaded regions represent a
  deviation within $\pm 1 \%$.
 }
\end{center}
\end{figure*}
Figure~\ref{F:fig_2pointf} shows a comparison of the 3D dark matter two-point
correlation function and power spectrum 
at $z=1.0$ and $z=0.06$ in the MBII and DMO simulations. To calculate the two-point correlation function, we randomly subsampled $4 \times 10^{5}$ dark matter particles and tested that the correlation function has already converged using a smaller subset. The power spectrum is calculated by taking the Fourier transform of the two-point correlation function,
\begin{equation} \label{eq:pspec}
P(k) = 4 \pi \int_{0}^{\sqrt{3}\Pi_\text{max}}\xi (r) r^{2} \frac{\sin{(kr)}}{kr}\,\mathrm{d}r.
\end{equation} 
From the left
panel of the figure, we
observe that $\xi(r)$ is larger at small scales ($<1\hmpc$) in the MBII
simulation, presumably due to baryonic physics. At these scales, the
correlation function is larger in MBII by a factor of $\sim 10 \%$,
and the discrepancy is larger 
at lower redshift. At intermediate scales (above $\sim 1\hmpc$), we find that the ratio of the correlation functions
approaches unity with no significant redshift dependence. The
enhancement of the correlation function at small scales is compensated
by a decrease in clustering at scales comparable to the box size.

 The
right panel of Fig.~\ref{F:fig_2pointf} shows a comparison of the
matter power spectra, demonstrating that 
the dark matter power spectrum in the hydrodynamic run is
enhanced by $\sim 10 \%$ at $k \gtrsim 10h/$Mpc, with the effect again
being stronger at low redshift. In contrast to our results, \cite{2011MNRAS.415.3649V} found that the dark
matter power spectrum is suppressed at intermediate scales in the OWLS simulation that included AGN feedback.  The purple band in this plot
indicates the target precision of $\sim$ 1 per cent accuracy in
predictions of the matter power spectrum
on scales of $0.1 h$ Mpc$^{-1} < k < 10 h$ Mpc$^{-1}$
\citep{2005APh....23..369H,2012JCAP...04..034H}. This level of accuracy in the theoretical
predictions is necessary to avoid systematic errors in constraining cosmological
parameters. \cite{2013PhRvD..87d3509Z,2014arXiv1405.7423E} and \cite{2014MNRAS.445.3382M} discuss approaches to
  mitigate the potentially large differences in the matter two-point correlations and reduce the
  uncertainty to the necessary level.

\subsection{Matching subhalos in MBII and DMO}
\begin{figure}
\begin{center}
\includegraphics[width=3.2in]{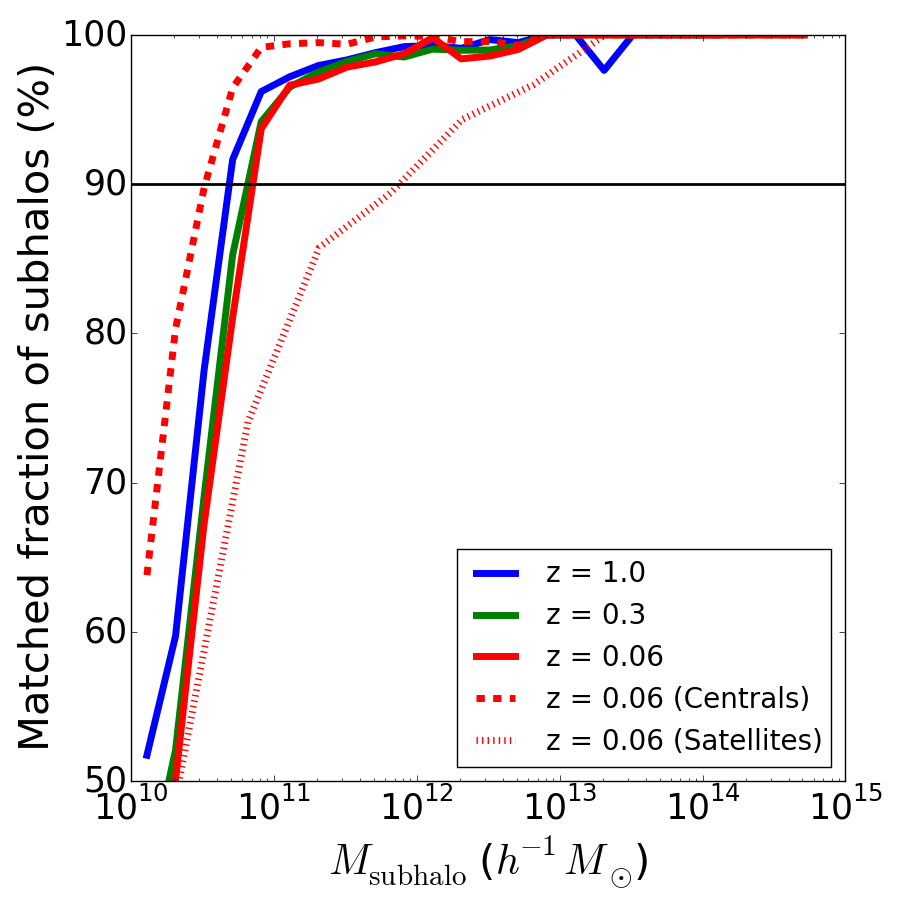}
\caption{\label{F:fig_sgmatchfrac} Fraction of subhalos matched at $z=1.0$, $0.3$, and $0.06$ as a
  function of the total subhalo mass in the MBII simulation.}
\end{center}
\end{figure}


To make a fair comparison between intrinsic alignments of galaxies and halos in the MBII
and DMO simulations, we matched the subhalos in both
simulations. 
The subhalos are matched using the unique ID's of the dark matter
particles in both the DMO and MBII simulations. For a given subhalo in
the MBII simulation with at least 1000 dark matter and star particles,
we identify the subhalo in the DMO simulation that has the highest number
of dark matter particles with the same IDs as in the MBII
subhalo. This subhalo in the DMO simulation is linked to the subhalo in
MBII if the fraction of common dark matter particles is greater than
$50 \%$ of the total number of dark matter particles in the subhalos
of each simulation. Figure~\ref{F:fig_sgmatchfrac} shows the matched fraction
of subhalos as a function of mass. We observe that the
fraction of matched subhalos increases with mass and approaches
unity. For the rest of this paper, 
we will only consider subhalos with mass greater than $10^{10.8}\hMsun$, where
the matched fraction exceeds $90 \%$.

The decrease in the fraction of matched subhalos at 
lower masses is due to an increase in the number of satellite
subhalos. The fraction of central subhalos matched even in the lower
mass bin, $10^{10.8-11.0}h^{-1}M_{\odot}$, is $\sim 99 \%$. As illustrated
in Figure~\ref{F:fig_halo0DM}, there are small but visible differences
in the smallest subhalo population in  a
matched halo of mass $\sim 10^{14}\hMsun$ (for example, some missing or merged
subhalos in MBII). We also attempted to match subhalos using only the
$50$ innermost dark matter particles, and obtained results that were
largely consistent with the method discussed earlier. However, for a
small fraction of subhalos, this new method would cause us to link
subhalos in the two simulations with very different masses. For
example, consider the distribution of dark matter in the MBII and DMO
simulations in Figure~\ref{F:fig_halo0DM}. The blue and green circles
show the virial radius of the central subhalo and a nearby satellite
subhalo respectively. We can clearly see that the center of the
satellite subhalo in MBII is closer to the center of the central
subhalo in DMO simulation, which leads to a false match using the $50$
closest particles. Hence, we use all the dark matter particles in the
subhalo to obtain a consistent match. For similar reasons (such as
changes in the location of density peaks), it is not possible to
consistently match the satellite subhalos in low mass halos, which is
the motivation for our adoption of a lower mass limit of $10^{10.8}h^{-1}M_{\odot}$.


\section{Shapes and misalignment angles} \label{S:radialshapes}
In this section, 
we investigate the change in the shape and orientation of the
subhalos with the distance from their center, which is defined as the
location of the most bound particle in the subhalo.  This measurement is necessary to
understand whether the shape of the stellar matter component traces the (inner)
shape of the dark matter component. We show the radial dependence of
the shape of the dark matter component in both the MBII and DMO
simulations, and compare the orientations at various radii against
that of the stellar component in MBII.

\subsection{Radial dependence of shapes} 
\begin{figure*}
\begin{center}
\includegraphics[width=2.25in]{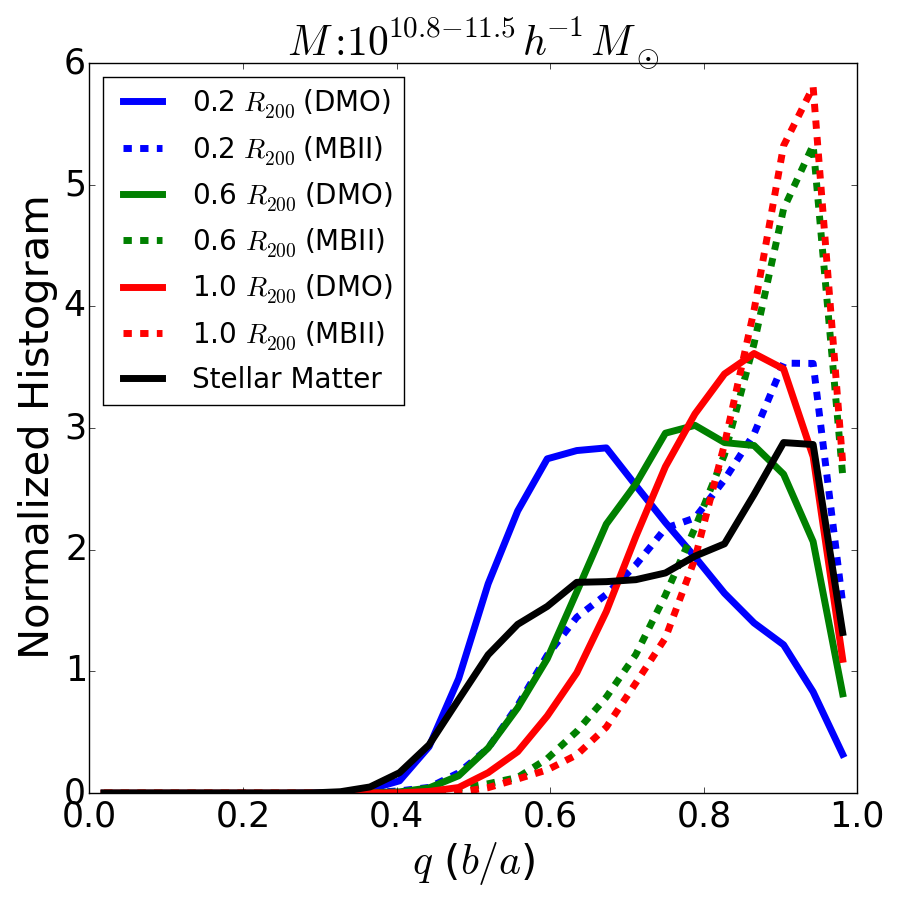}
\includegraphics[width=2.25in]{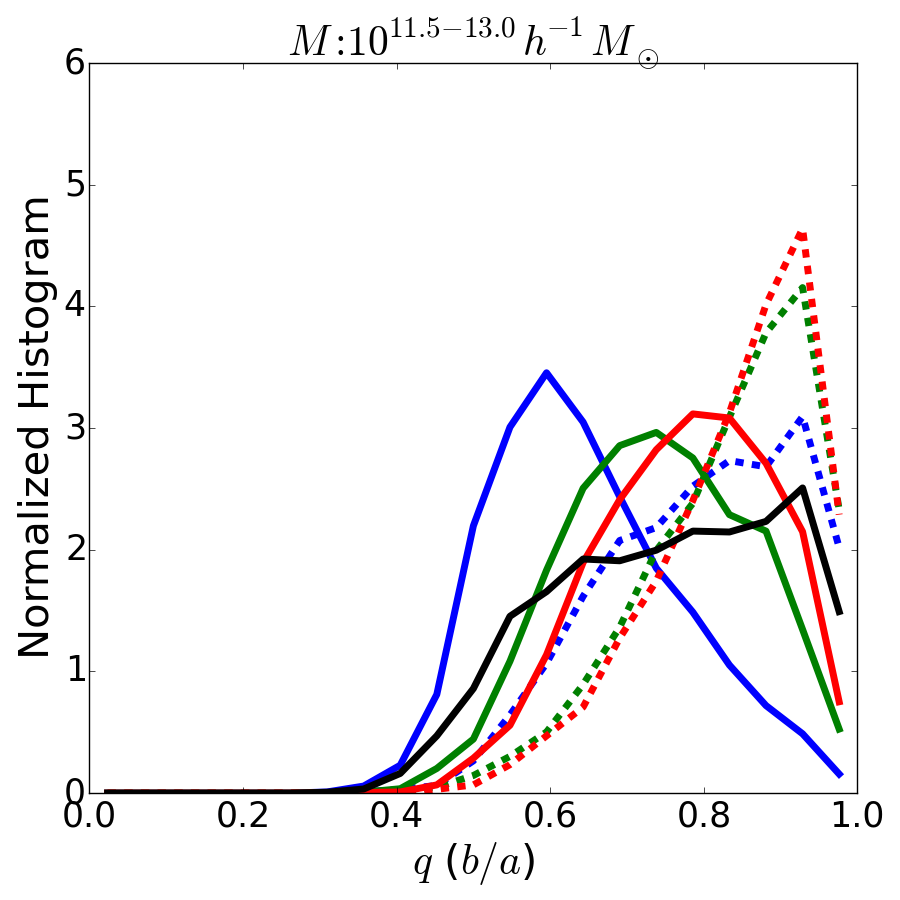}
\includegraphics[width=2.25in]{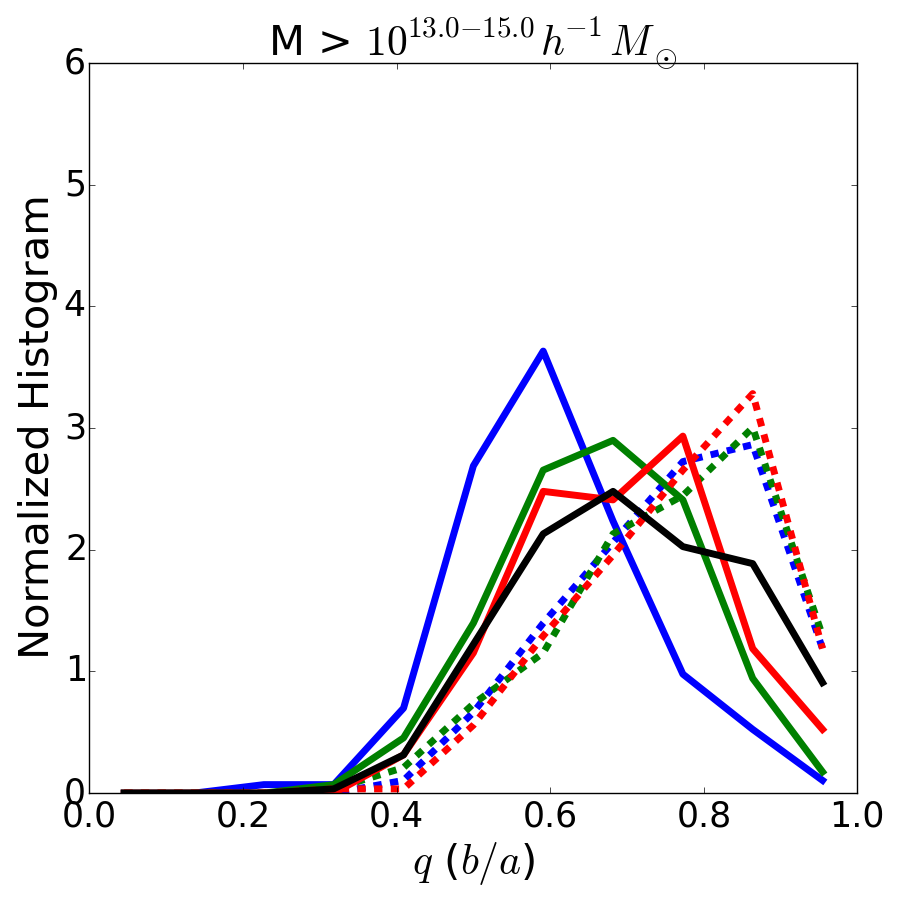}\\ 
\includegraphics[width=2.25in]{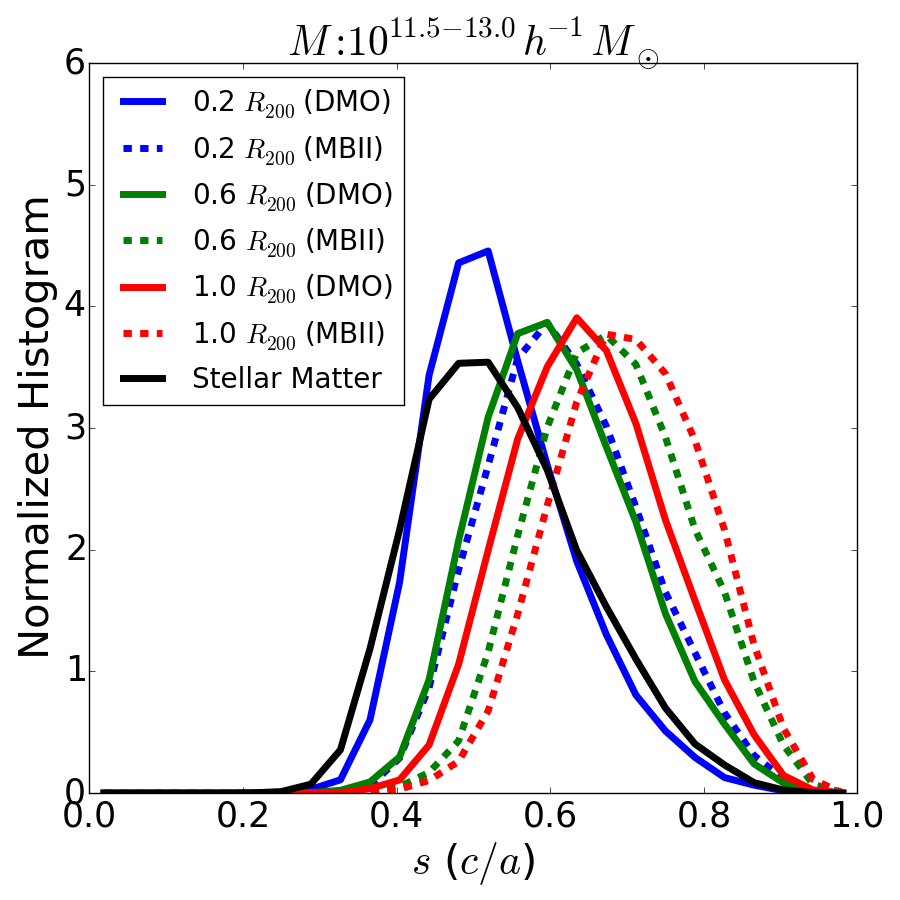}
\includegraphics[width=2.25in]{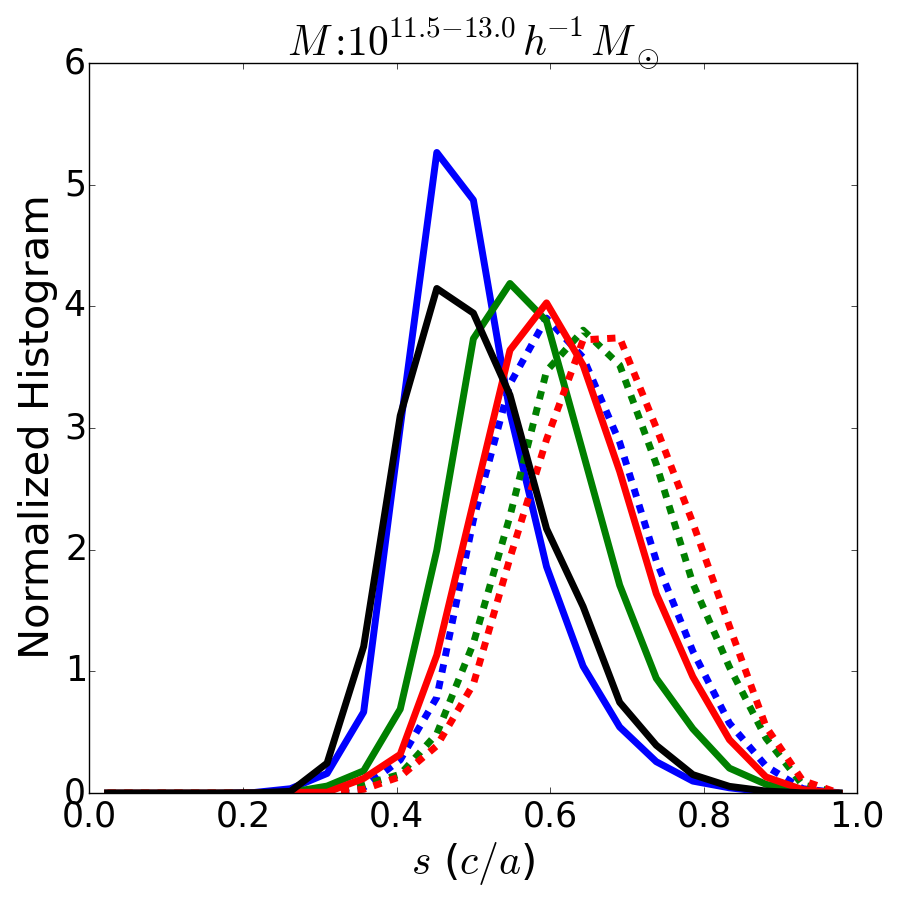}
\includegraphics[width=2.25in]{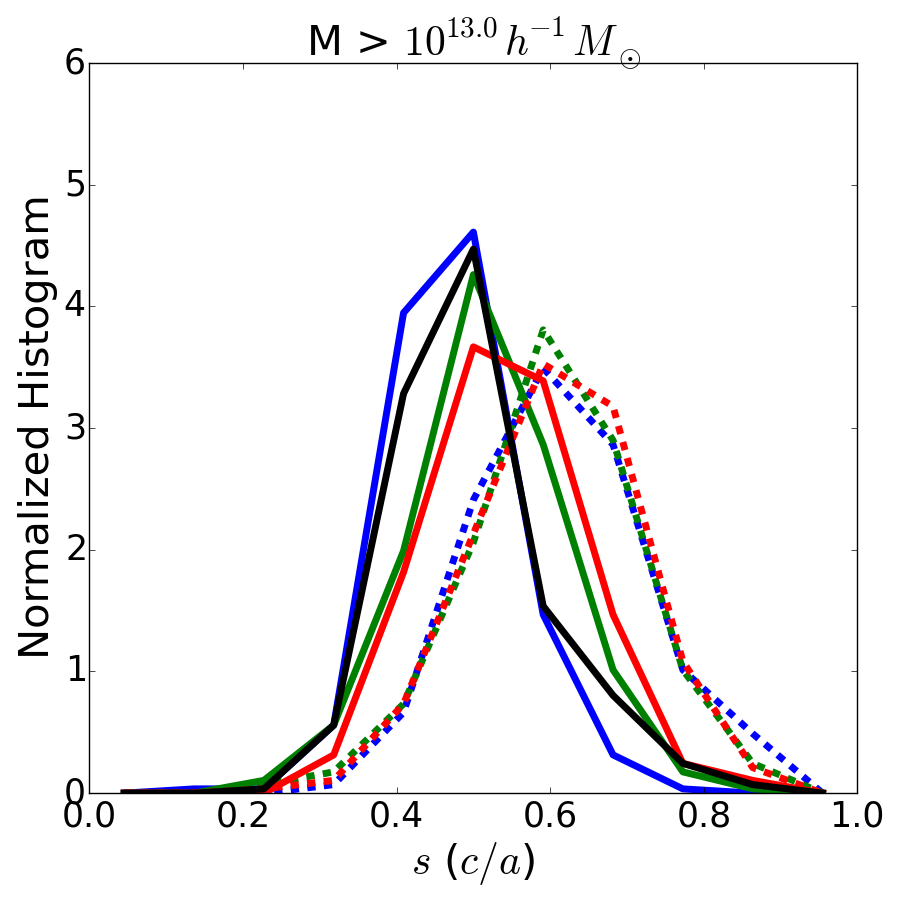}
\caption{\label{F:fig_mb2_qsradial} Normalized
  histograms of 3D axis ratios of the dark matter component in matched
  subhalos in the DMO and MBII simulations, with the shapes measured at
  different radii ($0.2R_{200}$, $0.6R_{200}$ and $1.0R_{200}$) and
  also the stellar matter component in MBII.  The columns indicate
  different mass bins, while the top and bottom rows are for $q~(b/a)$
  and $s~(c/a)$, respectively.}
\end{center}
\end{figure*}

\begin{figure*}
\begin{center}
\includegraphics[width=3.2in]{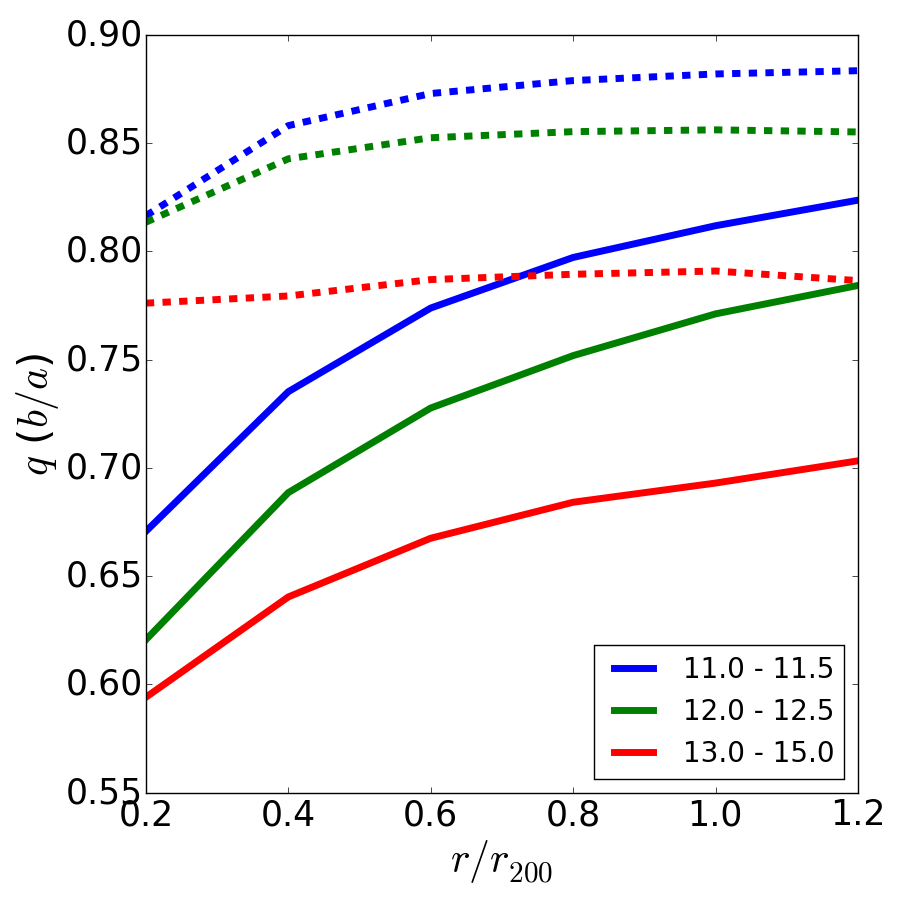}
\includegraphics[width=3.2in]{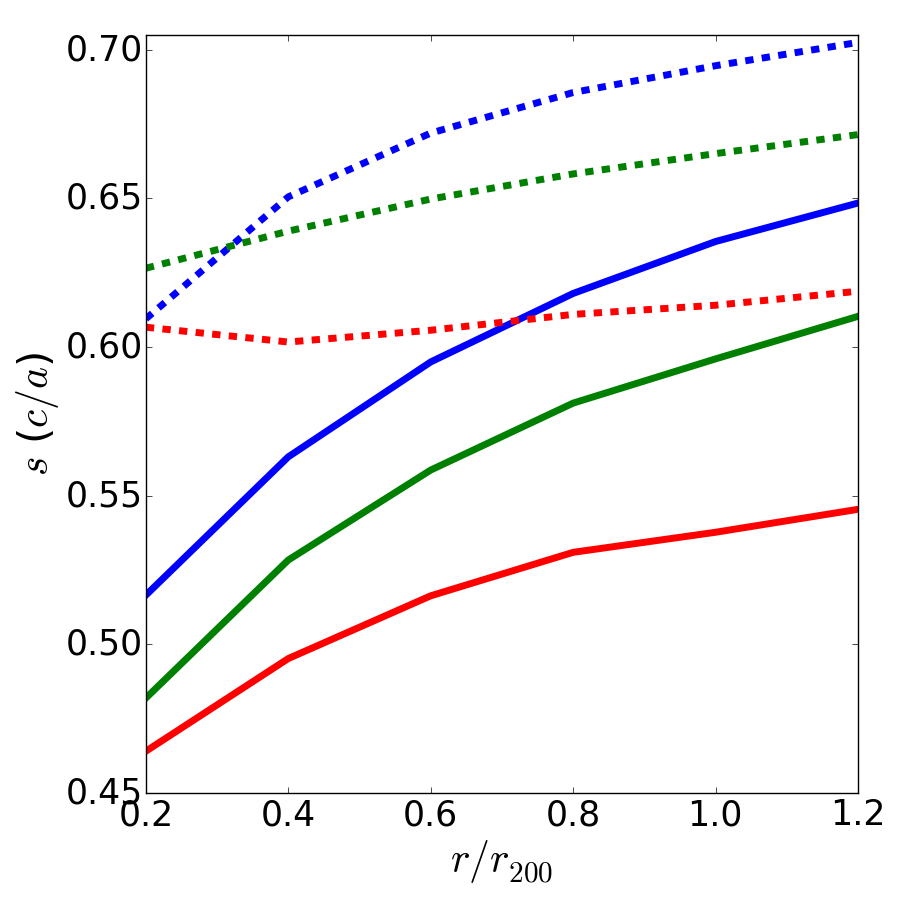}
\caption{\label{F:fig_dmomb2_qsmedian} Median of the dark matter
  subhalo axis ratios (left: $q$, right: $s$) in the MBII (dashed lines) and DMO (solid lines)
  simulations in different mass bins,
  plotted against the distance to which the shape is measured, at
  $z=0.06$.}
\end{center}
\end{figure*}

\begin{figure*}
\begin{center}
\includegraphics[width=2.25in]{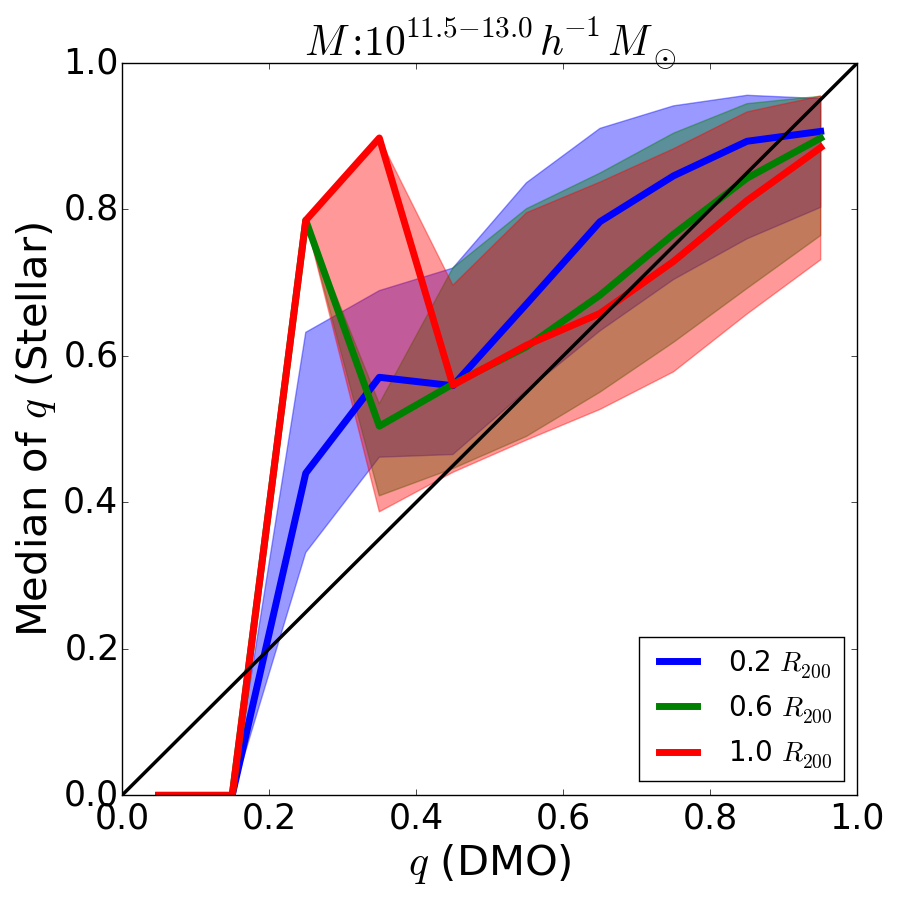}
\includegraphics[width=2.25in]{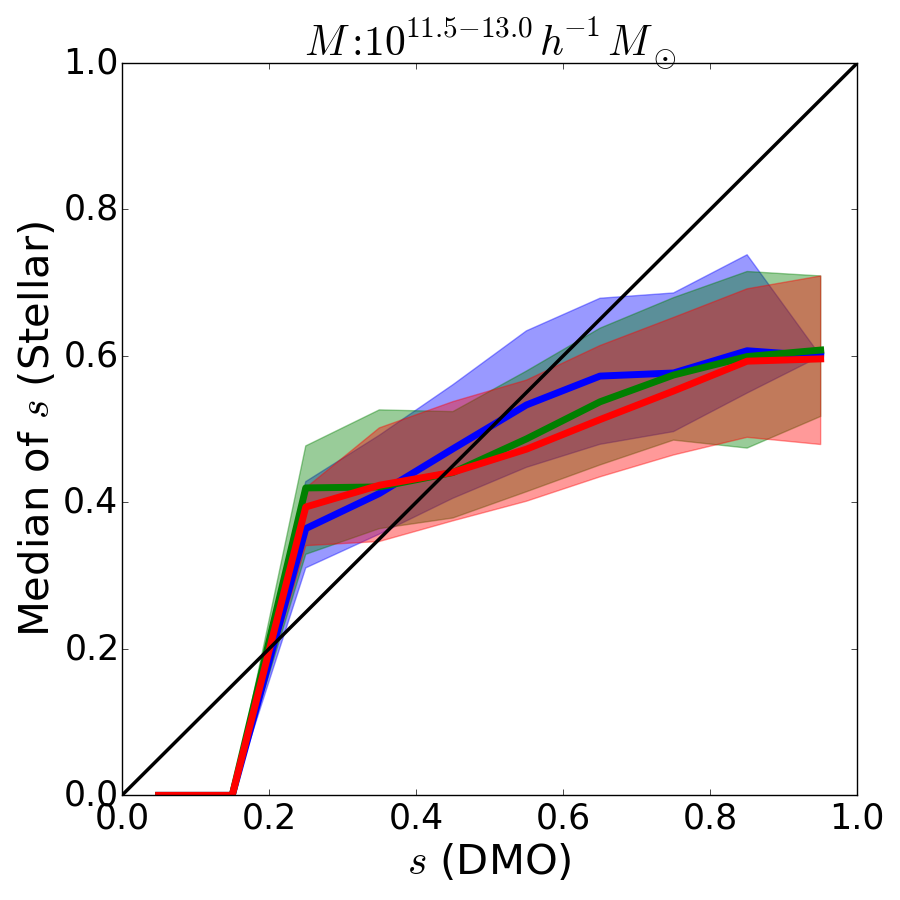}
\includegraphics[width=2.25in]{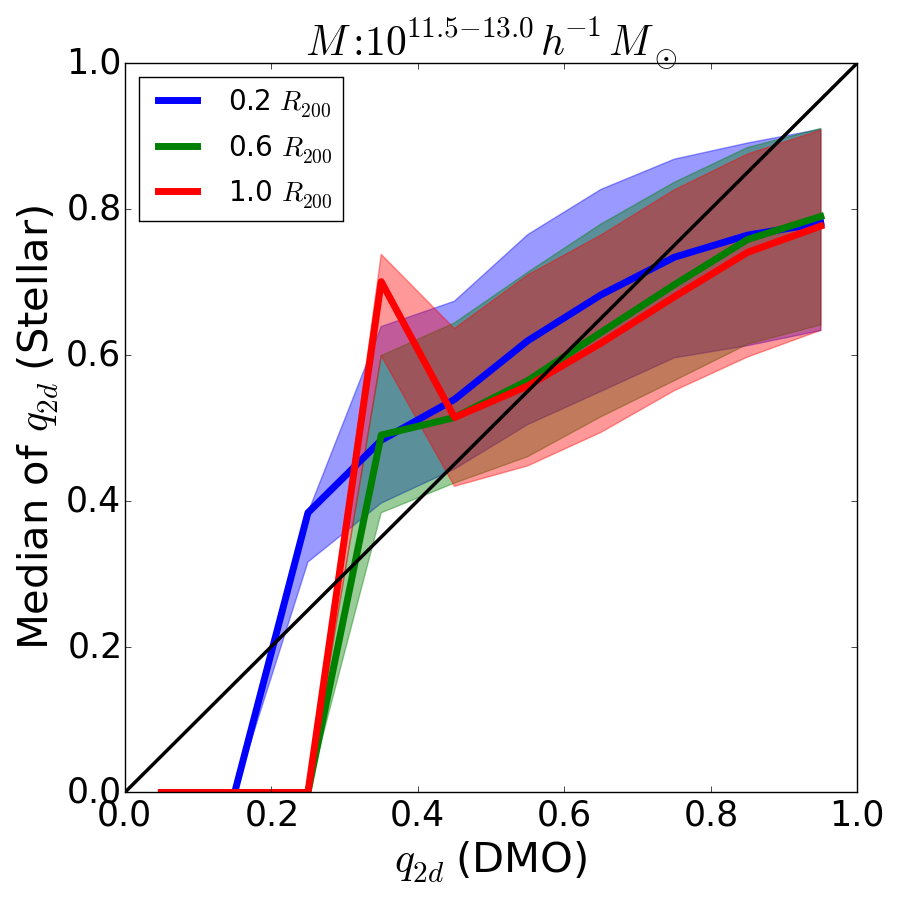}
\caption{\label{F:fig_dmo_qsradial} Median and scatter (defined using the $16^\text{th}$ and
  $84^\text{th}$ percentiles)  in the distribution of the axis
  ratios (3D and 2D) of the stellar matter component in MBII plotted against the
  axis ratio of the shape of matched subhalo in the DMO simulation, with the
  shape measured within different radii ($0.2R_{200}$, $0.6R_{200}$ and
  $1.0R_{200}$). {\em Left:} $q$ (3D); {\em middle:} $s$ (3D);
  {\em right:} $q_{2d}$ (2D).}
\end{center}
\end{figure*}

\begin{figure*}
\begin{center}
\includegraphics[width=2.25in]{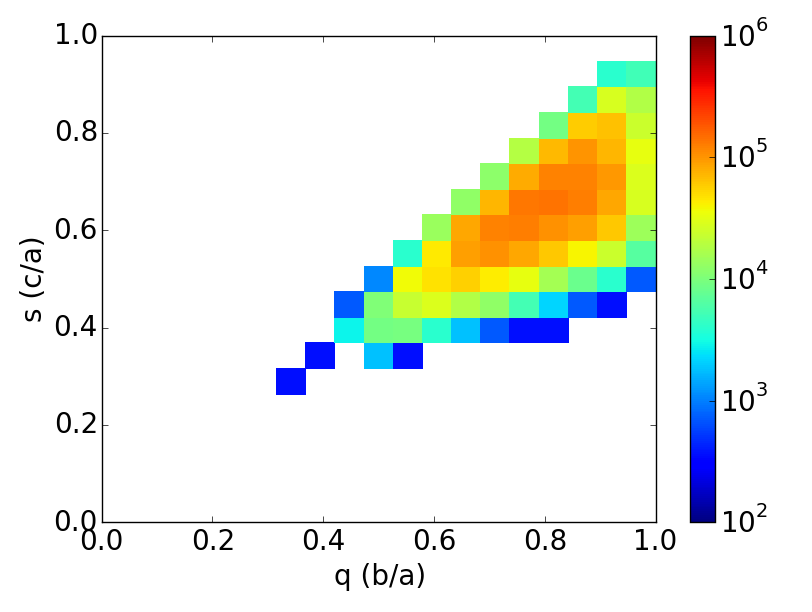}
\includegraphics[width=2.25in]{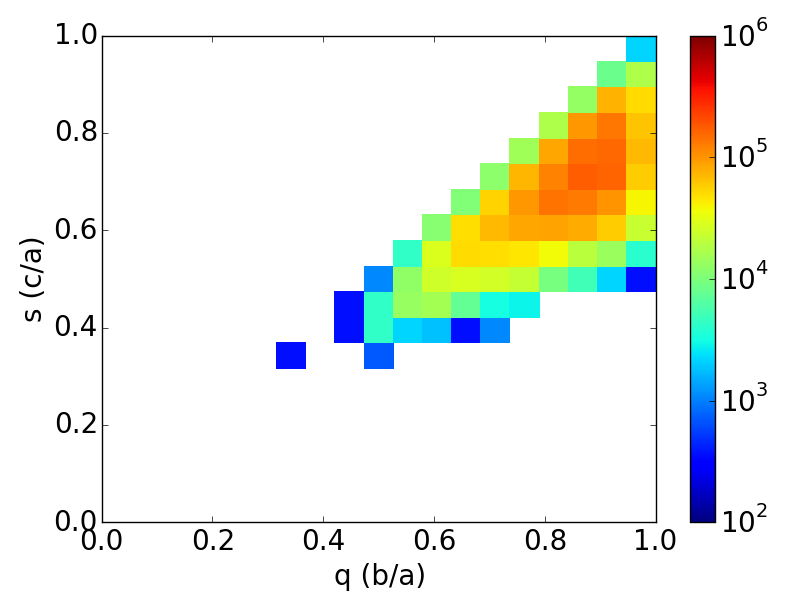}
\includegraphics[width=2.25in]{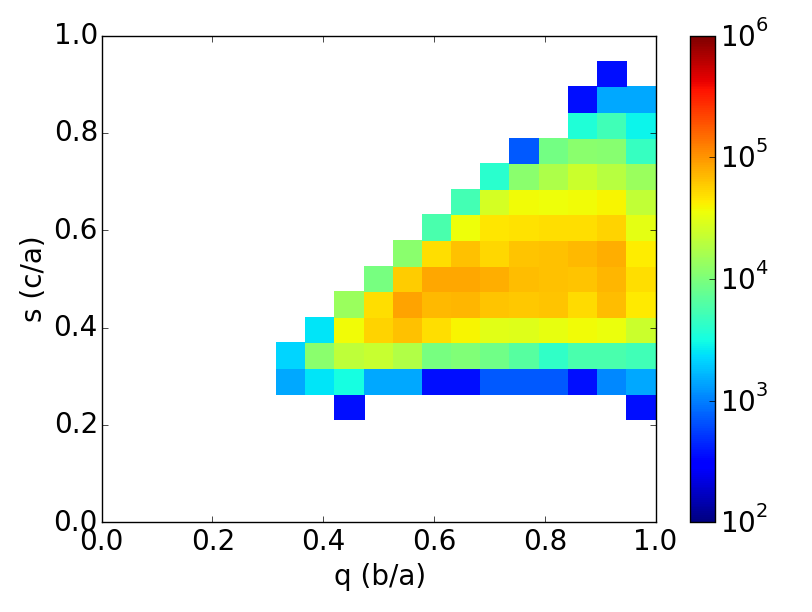}
\caption{\label{F:fig_qs_contour} Contour plots of 3D axis ratios of
  the dark matter component in matched subhalos in the DMO (left) and
  MBII (middle) simulations,  and the stellar matter component in MBII
  (right), in the mass bin M2  at  $z=0.06$.}
\end{center}
\end{figure*}

To calculate the shape of the dark matter component in the subhalo within a given radius, we start with all
dark matter particles inside the spherical volume at a given distance from the center and compute
the shape using the iterative reduced inertia tensor
(Eq.~\ref{eq:redinertensor}). By using the reduced inertia tensor for
shape calculation, we still provide more weight to the particles in
the inner region of the subhalo, so our resulting shapes should be
considered as shapes within a radius rather than at that
radius. Hence, the effect of baryonic physics on the dark matter halo
shapes in MBII will be evident even for shapes measured out to large radii. For the analysis of shapes, we only consider subhalos with at least
500 particles remaining after the last iteration, excluding fewer than
$1 \%$ of galaxies. We will compare the distribution of
shapes and alignments of the dark matter component calculated within a radial distance of $0.2 R_{200}$,
$0.6 R_{200}$ and $1.0 R_{200}$. Here $R_{200}$ is the radius within
which the spherically-calculated average density of matter is equal to
$200$ times the critical density.

 The radial dependence of dark matter
halo shapes has previously been studied using $N$-body simulations
\citep[e.g.,][]{2006MNRAS.367.1781A,2012JCAP...05..030S} and small-volume
hydrodynamic simulations
\citep[e.g.,][]{2004ApJ...611L..73K,2010MNRAS.407..435A,2011MNRAS.415.2607D}. As
in our work, the
method used to measure the shapes in \cite{2006MNRAS.367.1781A},
\cite{2011MNRAS.415.2607D}, and \cite{2012JCAP...05..030S} is based on
the iterative reduced inertia tensor. However,
\citep{2004ApJ...611L..73K} measure shapes using only particles within  different radial bins 
with the reduced inertia tensor, and \cite{2010MNRAS.407..435A} fit
ellipsoids to the positions of dark matter particles along the
isopotential contours at different radii. Hence, the comparison with
some of these previous studies is only qualitative.    

In Figure~\ref{F:fig_mb2_qsradial}, we show the normalized histograms
of the dark matter subhalo axis ratios 
in MBII and DMO calculated with different maximum radii,
as well as the axis ratios of the total stellar matter component of
MBII.  Throughout this section, we use three mass bins: $10^{10.8-11.5}\hMsun$,
$10^{11.5-13.0}\hMsun$ and $>10^{13}\hMsun$ (M1, M2, and
M3). Comparing the dark matter subhalo axis ratio distributions in MBII and DMO
simulations, we observe that 
for a given maximum radius, the shapes measured in the hydrodynamic run are more spherical. This
finding  is in agreement with results from previous studies using smaller-volume hydrodynamic simulations
  \citep{2004ApJ...611L..73K,2010MNRAS.407..435A,2011MNRAS.415.2607D}. We also find that the axis
ratios increase as we go to larger radii, which means that the shape
of the dark matter component is more 
flattened in the inner regions of the subhalo, again in agreement with previous findings
\citep{2011MNRAS.415.2607D,2012JCAP...05..030S}. 

To illustrate this effect further, we plot the median 
axis ratios against the radius within which the shape is measured in
Figure~\ref{F:fig_dmomb2_qsmedian}, in more narrowly-defined mass bins. From the plot, we can see
that for a given mass bin, the median axis ratios are higher in MBII. The median
values of $q$ and $s$ can be compared against those from \cite{2012JCAP...05..030S}, and 
qualitatively we reproduce their trend that they decrease at small 
distances from the subhalo center and at increasing mass.  However, this increase in axis
ratios is smaller in MB-II than in the DMO simulation. Also, at higher
masses, the increase in the median axis ratio with radius is milder
than at lower masses. Using smaller-volume hydrodynamical simulations, \cite{2010MNRAS.407..435A}
found that the halo axis ratios are independent of radius, and \cite{2011MNRAS.415.2607D} found that the
shapes of dark matter halos are slightly more oblate in the inner regions. These differences
are most likely due to the absence of stellar and AGN feedback in
those studies, unlike in MBII.    

From Figure~\ref{F:fig_mb2_qsradial}, we see that the axis ratio
histograms for the dark matter subhalos in the DMO simulation do not
follow those for the galaxy stellar components in MBII, at any radius. 
 This result can be seen more clearly in
 Figure~\ref{F:fig_dmo_qsradial}, where we plot the median and scatter in the
 stellar matter axis ratio distributions in MBII as a function of the
 DMO dark matter $q$ or $s$ value, for different radii. We see that the scatter in the distribution of axis
 ratios is large for all radii, so there is no advantage in using the inner shape
 of the dark matter subhalo in dark matter-only simulations to predict
 the shape of the stellar component in
 MBII. For this reason,  we only consider the dark matter subhalo axis
 ratios using all particles in our 
analysis of two-point statistics.

 In Figure~\ref{F:fig_qs_contour}, we show the contour plot of $q$ versus $s$ for the dark matter
 shape in DMO and MBII and stellar shape in MBII in the mass bin M2 
 ($10^{11.5-13.0}\hMsun$). 
 The contour plots indicate that the galaxy shapes are more
 prolate compared to the shapes of the dark matter component in MBII. We can use the
 triaxiality parameter, $T = \frac{1 - q^{2}}{1 - s^{2}}$ \citep{1991ApJ...383..112F}, to quantify the prolateness or oblateness
 of the shape. Large (small) values of $T$ imply that the shape is
 more prolate (oblate). In the mass bin M2, the mean triaxiality of
 the stellar component shapes is $0.562 \pm 0.003$, while for dark
 matter shapes in MBII, the mean triaxiality is $0.538 \pm 0.002$. The
 triaxialities are larger for the dark matter shapes in the DMO
 simulation with a mean value of $0.600 \pm 0.002$, meaning that the
 dark matter shapes are more oblate (prolate) in hydrodynamic (dark matter-only) simulations. This conclusion is consistent with results from previous comparisons performed using simulations of smaller volume \citep{2006ApJ...648..807B,2006PhRvD..74l3522G,2010MNRAS.407..435A,2011MNRAS.415.2607D}.

\subsection{Misalignment angles}
\begin{figure*}
\begin{center}
\includegraphics[width=2.25in]{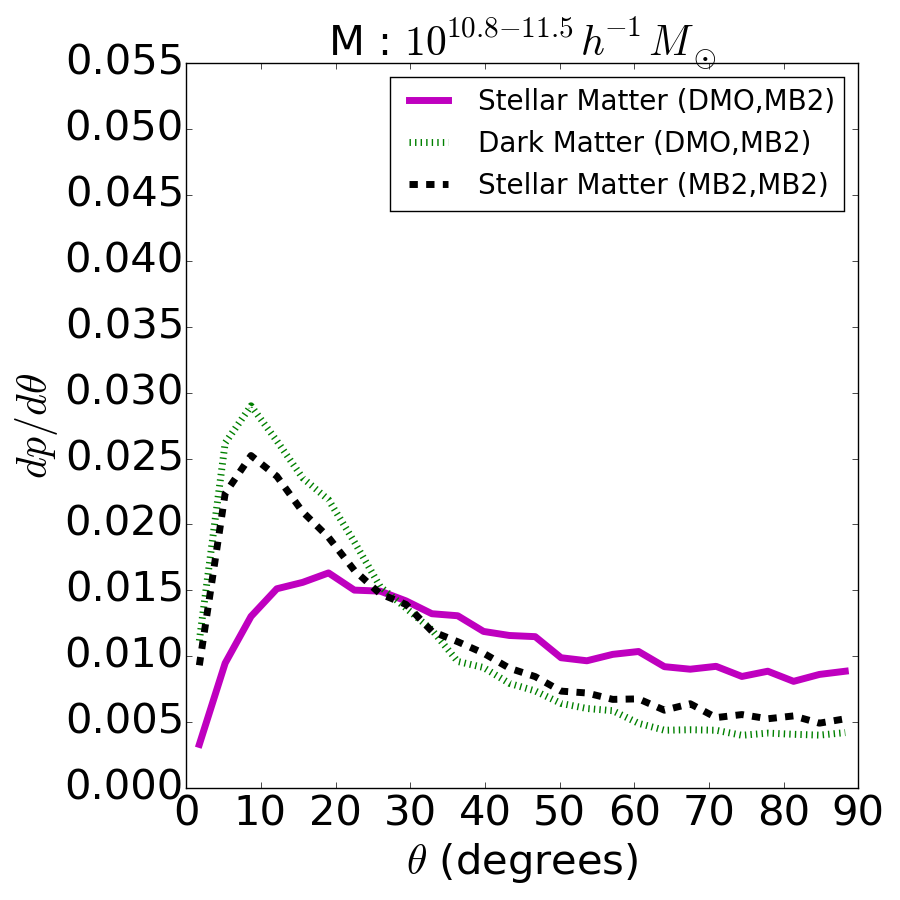}
\includegraphics[width=2.25in]{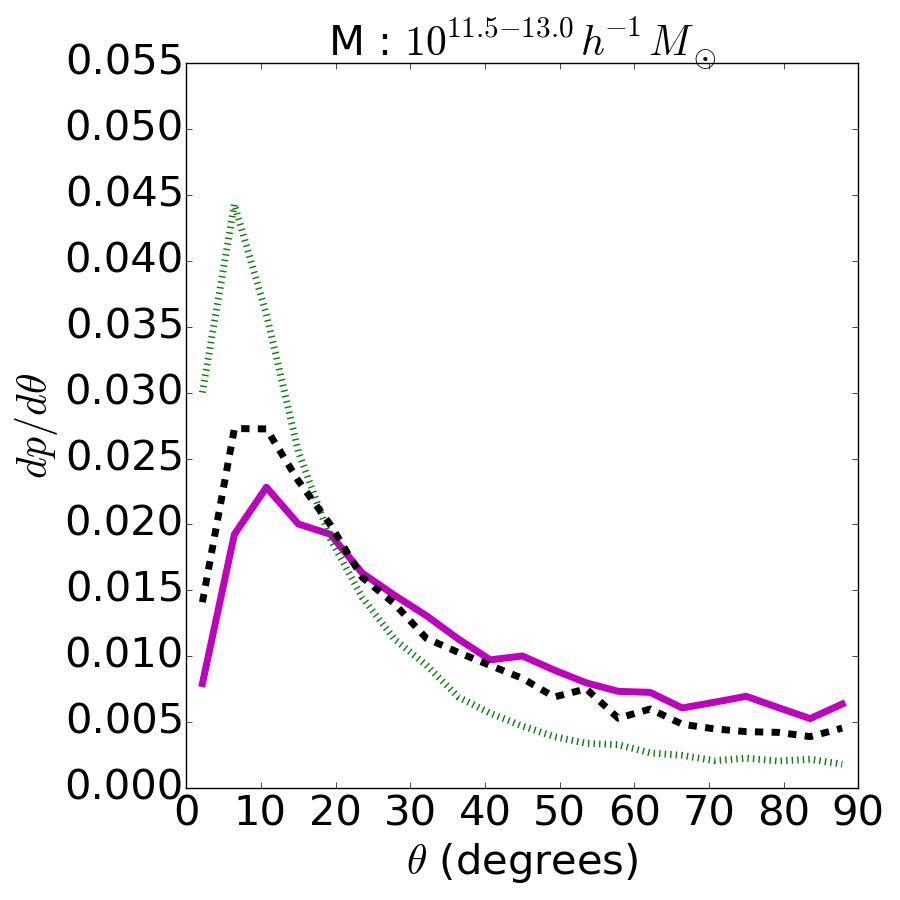}
\includegraphics[width=2.25in]{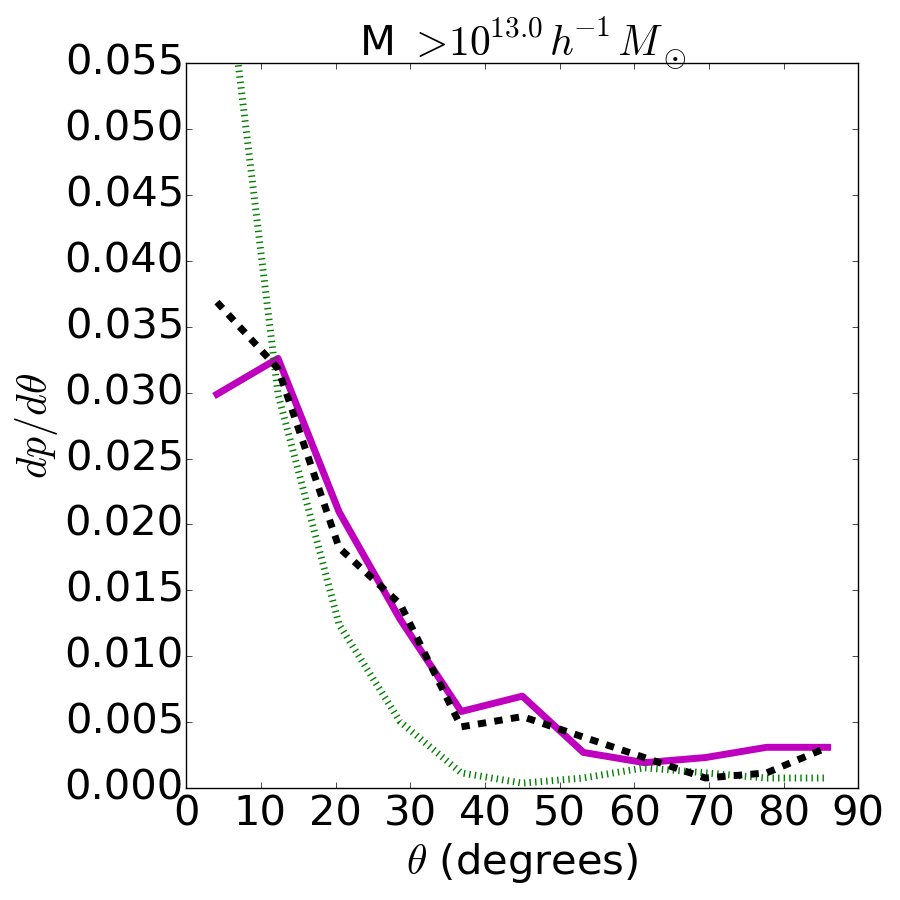}
\caption{\label{F:fig_dm_ma3d} Misalignment angle distributions for
  the 3D shapes of the dark matter component in matched subhalos of
  DMO and MBII simulations with the stellar matter component in MBII, in  mass bins M1, M2 and M3 at
  $z=0.06$. Also shown (green line) is the histogram of misalignment angles between the shapes of dark matter subhalos in MBII and DMO.}
\end{center}
\end{figure*}

\begin{figure*}
\begin{center}
\includegraphics[width=2.25in]{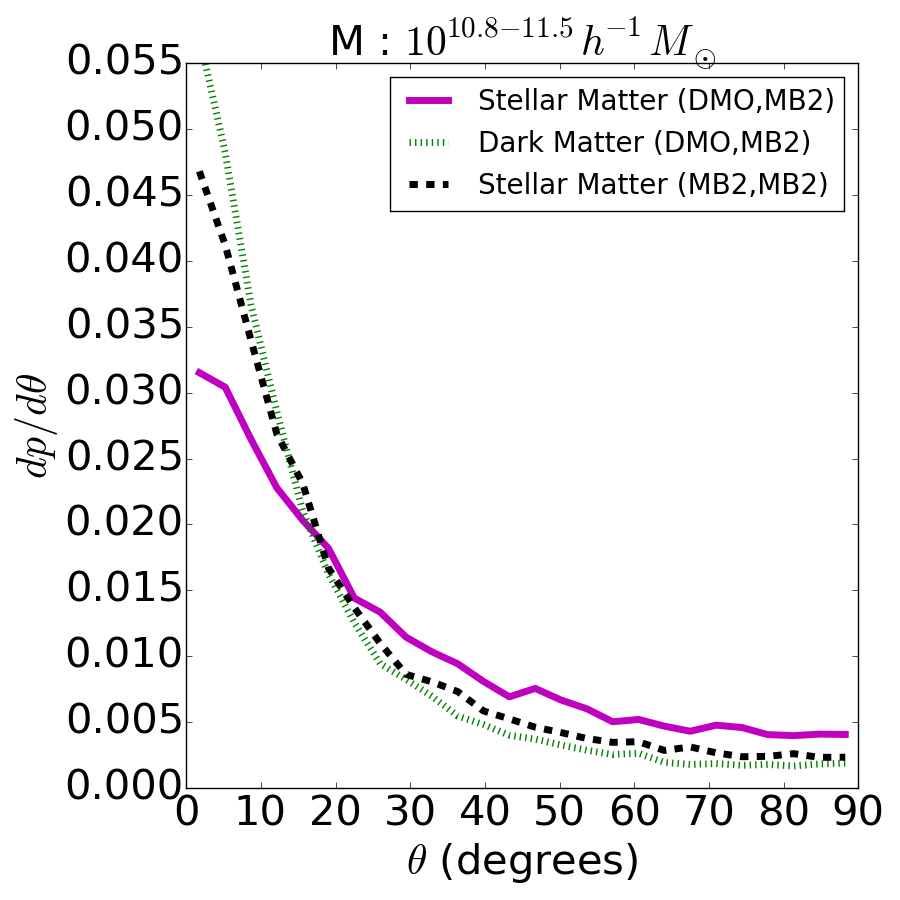}
\includegraphics[width=2.25in]{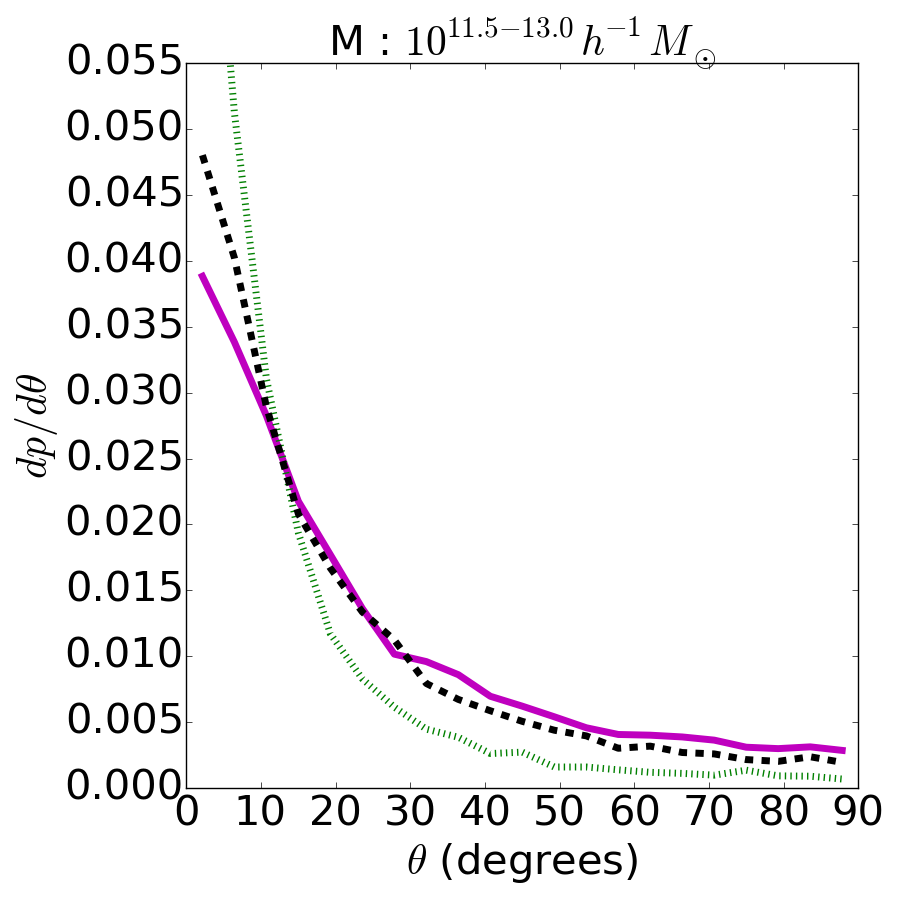}
\includegraphics[width=2.25in]{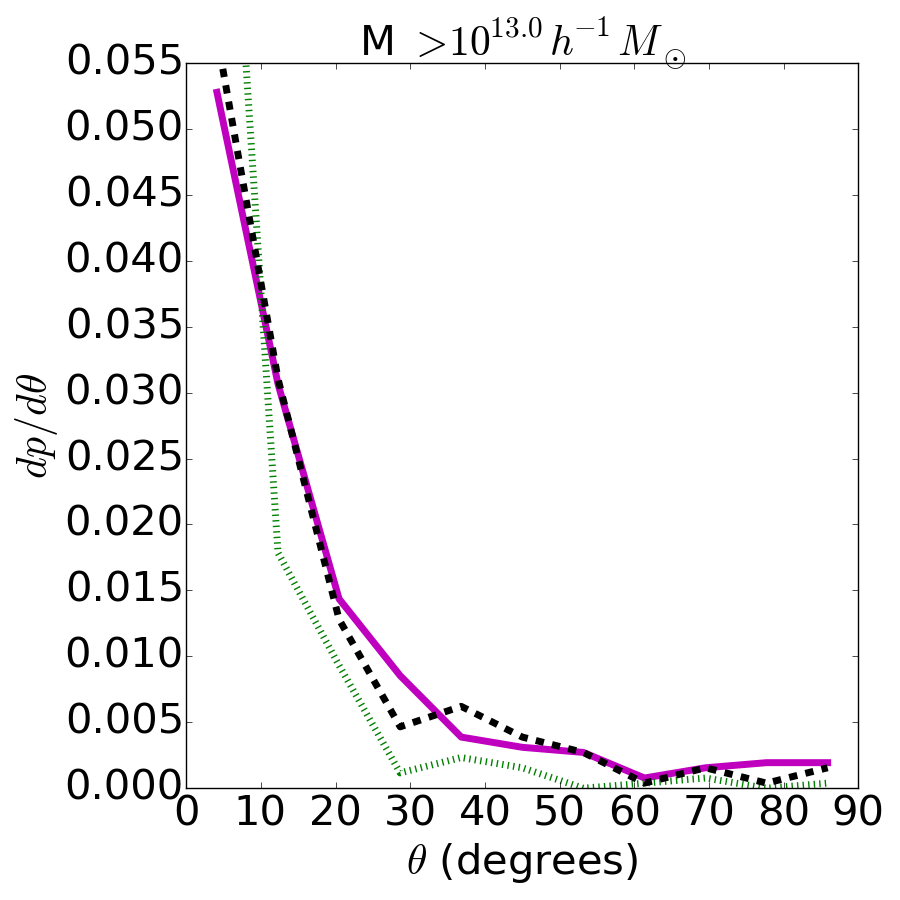}
\caption{\label{F:fig_dm_ma2d} Misalignment angle distributions for 2D
  shapes of the dark matter component in matched subhalos of DMO and
  MBII with the stellar matter component in MBII, in mass bins M1, M2 and M3 at
  $z=0.06$. Also shown (green line) is the histogram of misalignment angles between the shapes of
  dark matter subhalos in MBII and DMO.  This figure is simply the 2D version of Fig.~\ref{F:fig_dm_ma3d}.}
\end{center}
\end{figure*}

\begin{figure}
\begin{center}
\includegraphics[width=3.2in]{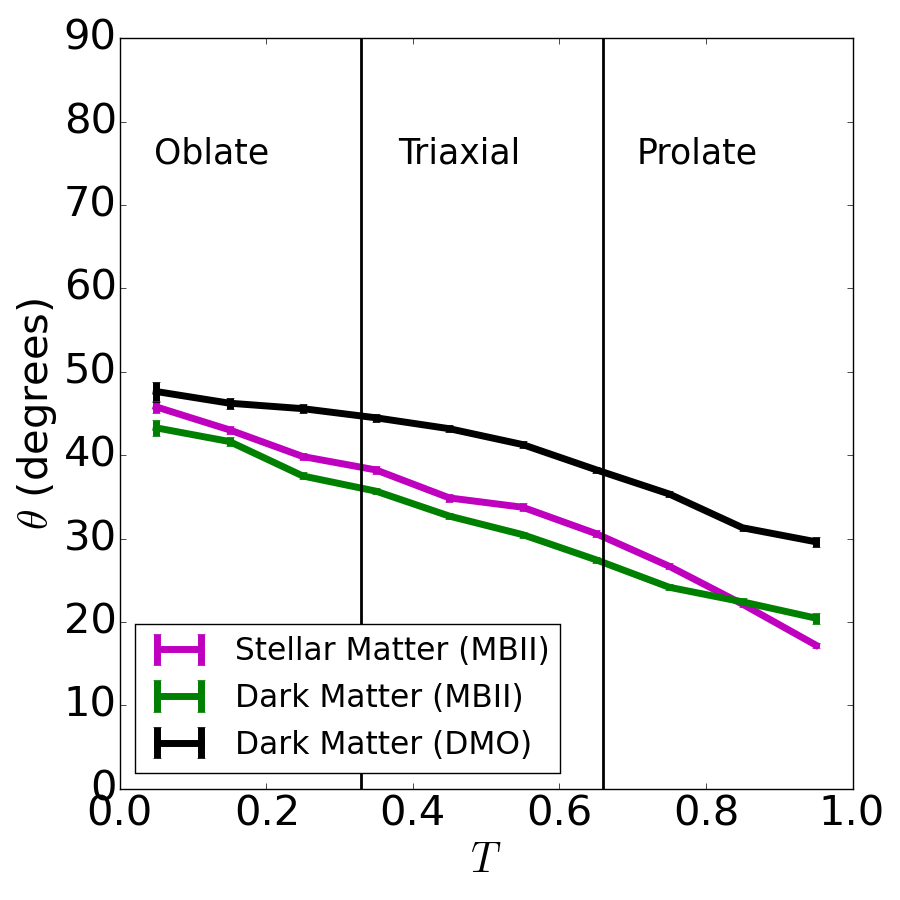}
\caption{\label{F:fig_Tma} Mean of the 3D misalignment angles between the
  shape of the dark matter component in MBII and DMO with the shape of
  stellar component in MBII as a function of the triaxiality
  parameter, $T$ (of the stellar matter component in MBII and the dark matter component in MBII and DMO), in the mass range, $10^{10.8} - 6.0 \times 10^{14} \hMsun$ at $z=0.06$. The purple line shows the mean misalignment angle between the shape of dark matter component in MBII with the stellar component in MBII plotted against the triaxiality of the shape of stellar component in MBII. Similarly, the green and black lines show the mean misalignment angles between the shapes of dark matter component in MBII and DMO with the stellar component in MBII plotted against the triaxialities of the shapes of dark matter component in MBII and DMO respectively.
}
\end{center}
\end{figure}
In Fig.~\ref{F:fig_dm_ma3d}, we show the normalized histograms of the misalignment angles
(Eq.~\ref{eq:misalignangle}) between the 3D shape defined by the stellar component of subhalos in
MBII and the shape defined by the dark matter component in MBII and DMO (for mass bins M1, M2 and
M3). From the plot, we can see that for M3, the alignment of stellar matter in MBII with the dark
matter component in DMO (purple curve) is similar to the alignment with the dark matter component in
MBII (dashed black curve). This result implies that the shapes of dark matter components in the
matched subhalos of MBII and DMO have similar orientations at high mass, which is also shown
directly by the green line. Fig.~\ref{F:fig_dm_ma2d} shows a similar result using the projected
shapes, with slightly smaller misalignment angles.  However, for M1 and M2, it is clear that the
stellar matter in MBII is better aligned with the MBII dark matter subhalo than with
the corresponding subhalo in DMO.

To check for a connection between galaxy shapes and misalignment angles, we
consider the triaxiality parameter, $T$. In
Figure~\ref{F:fig_Tma}, we plot the mean misalignment angles between the
shapes of stellar matter in MBII with the dark matter shapes in DMO and MBII
simulations as a function of triaxiality. From the
figure, we can see that as $T$ increases for the stellar shapes and dark matter
shapes in MBII, the mean misalignment angles decrease. This means that for
stellar and dark matter shapes in MBII which are more prolate, the alignment
between the shapes of stellar and the dark matter components is stronger than for those with 
more oblate shapes. Similarly, for the more prolate dark matter shapes in
the DMO simulation, the alignment with the stellar
component in MBII is closer. However, the mean misalignment angles decreases
by only $\sim 27\%$ while going from  $T < 0.33$ to $T > 0.66$ in DMO, which is less than the decrease of $\sim 45 \%$ for stellar
shapes in MBII. When using galaxy shapes, it is tempting to interpret more oblate shapes as relating
to disk galaxies, and thus inferring that disk galaxy shapes are more misaligned with their host
dark matter halos than elliptical galaxies.  However, we defer a more detailed comparison of the
morphologies of galaxies and the 
connection to  misalignment angle distributions in future work. 
        
\subsection{Radial dependence of misalignment angles}
\begin{figure*}
\begin{center}
\includegraphics[width=2.25in]{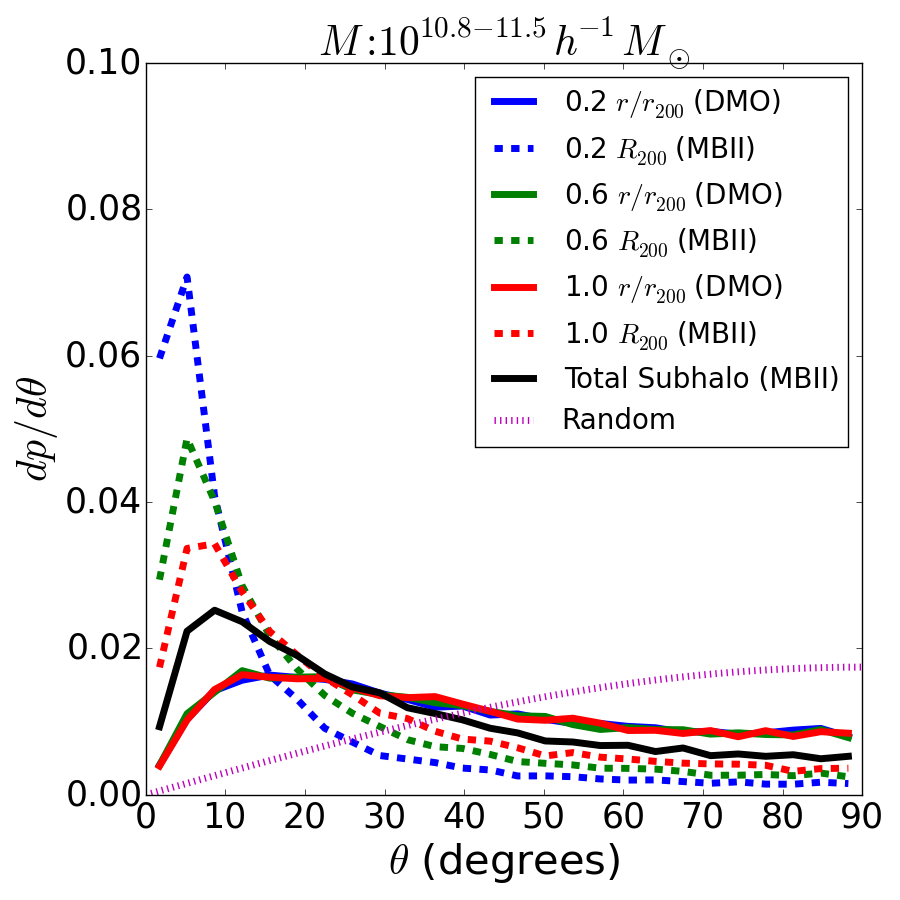}
\includegraphics[width=2.25in]{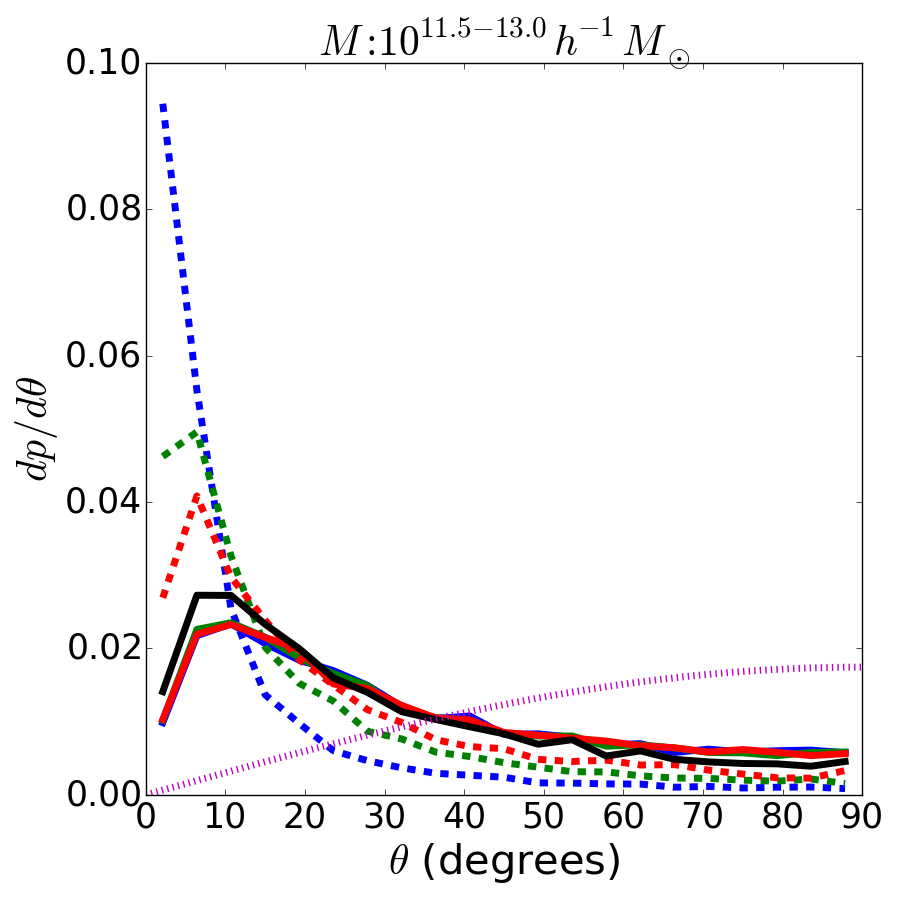}
\includegraphics[width=2.25in]{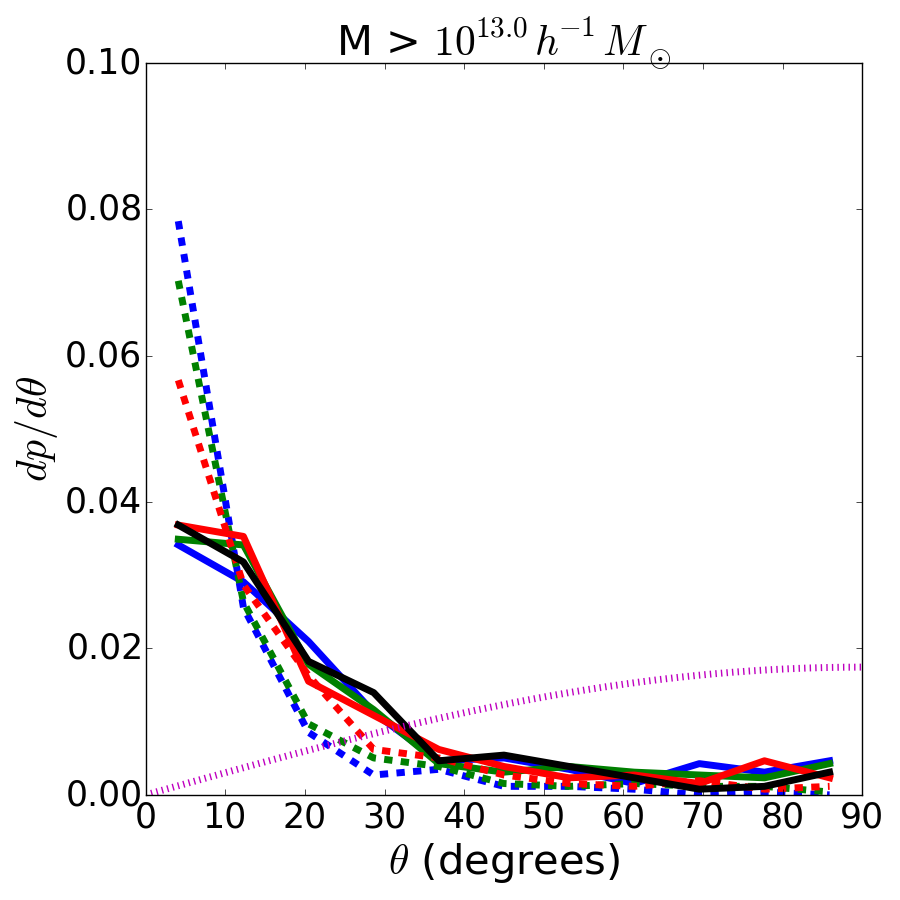}
\caption{\label{F:fig_mb2_maradial} Histograms of misalignment angles of 3D shapes of dark matter
  subhalos in the DMO and MBII simulations (with respect to the shape of the galaxy in MBII) when
  measuring the subhalo shapes within different
  radii ($0.2R_{200}$, $0.6R_{200}$, and $1.0R_{200}$)
  in the  mass bins M1 (left), M2 (middle), and M3 (right) at $z=0.06$.}
\end{center}
\end{figure*} 

Next, we investigate radial trends in the orientation of the shape of the stellar component in MBII
with respect to the dark matter component at different radii in MBII and DMO.  Histograms of the
misalignment angles when defining the dark matter halo shape within various radii are shown in
Figure~\ref{F:fig_mb2_maradial}. For comparison, we also show the prediction for a
purely random distribution of misalignment angles in 3D. 
From the plots, we see that in the MBII simulation, the alignment of the stellar component with the
dark matter subhalo shape increases as we limit ourselves to smaller radii within the dark matter
subhalo. This is expected, as the stellar matter is coupled to the distribution and shape of the
dark matter in the inner regions of the subhalo. However, there is no corresponding trend in the DMO
simulation, where the misalignment angle distributions have very little dependence on the maximum radius.

From this plot, we can conclude that it is not advantageous to use the inner shape of the dark
matter subhalo in a dark matter-only simulation when trying to define mock galaxy shapes and
alignments.  Since the distribution of misalignment angles with respect to the total dark matter
subhalo orientation is not consistent with a uniform random distribution, we can still use these distributions in different mass bins to assign shapes and orientations to galaxies painted onto $N$-body simulations.

\section{Intrinsic alignment two-point correlation functions}\label{S:edwdp}
\begin{figure*}
\begin{center}
\includegraphics[width=2.25in]{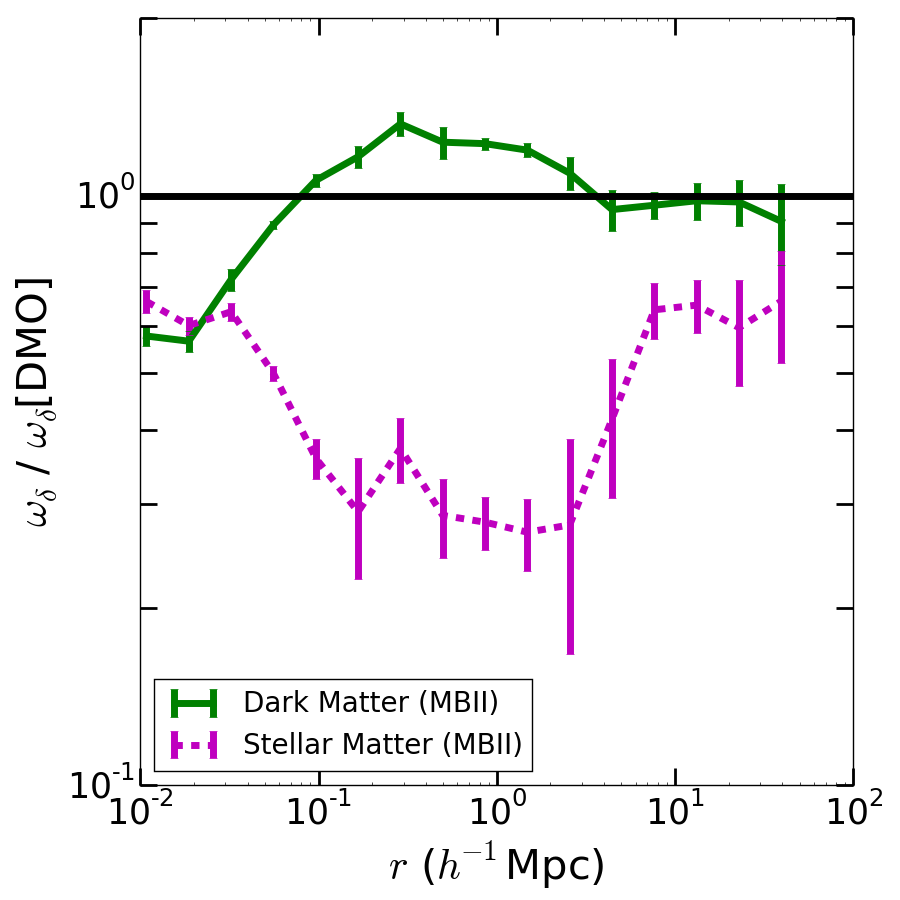}
\includegraphics[width=2.25in]{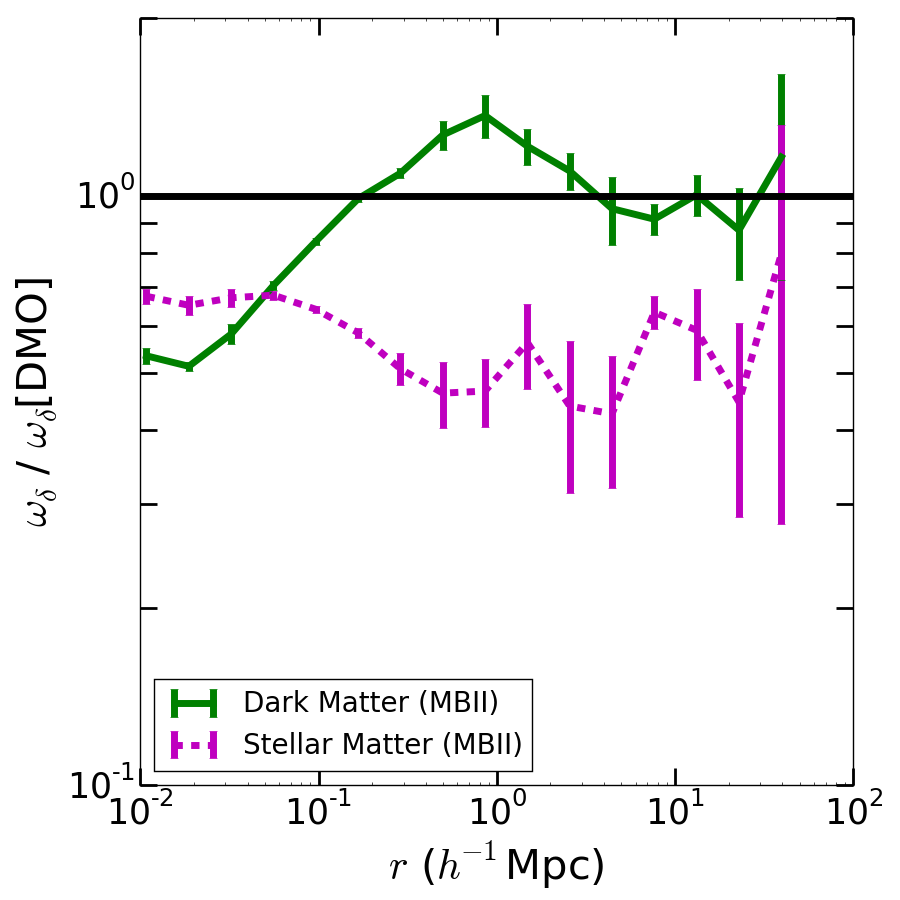}
\includegraphics[width=2.25in]{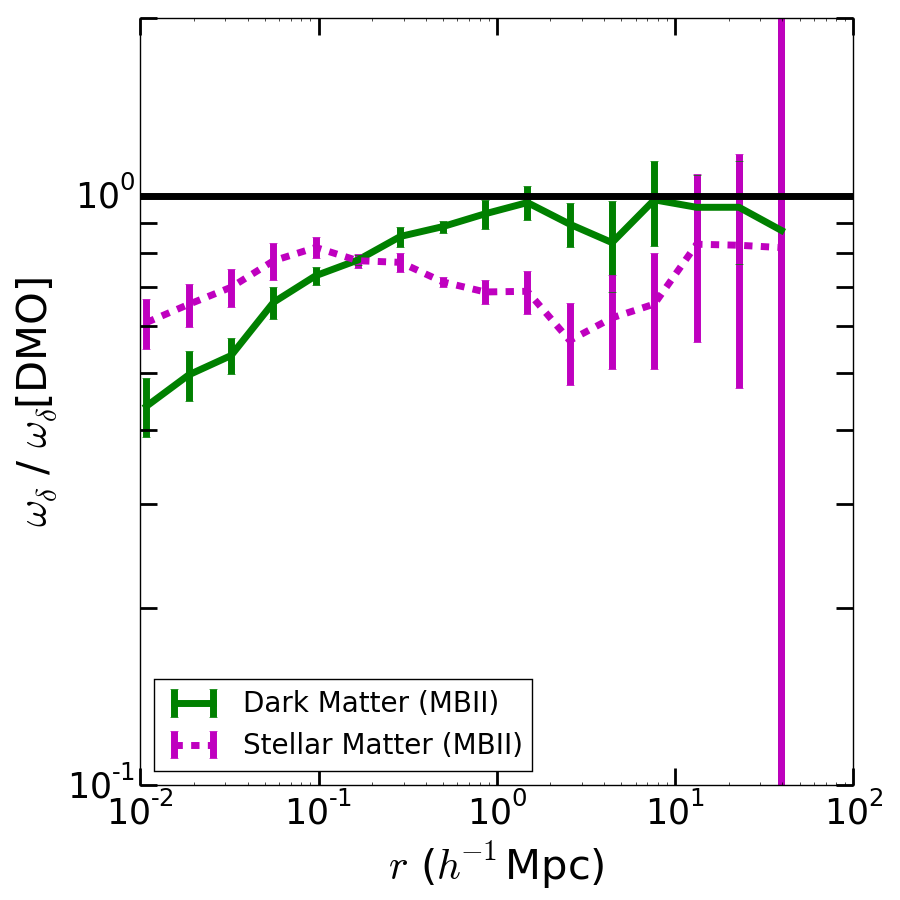}
\caption{\label{F:fig_ed} Comparison of the ED correlation function, $\omega_{\delta}(r)$, for the
  dark matter subhalo and stellar matter components in MBII with respect to that for dark matter
  subhalos in the DMO simulation, computed separately for M1 (left), M2 (midddle), and M3 (right).}
\end{center}
\end{figure*}

\begin{figure*}
\begin{center}
\includegraphics[width=2.25in]{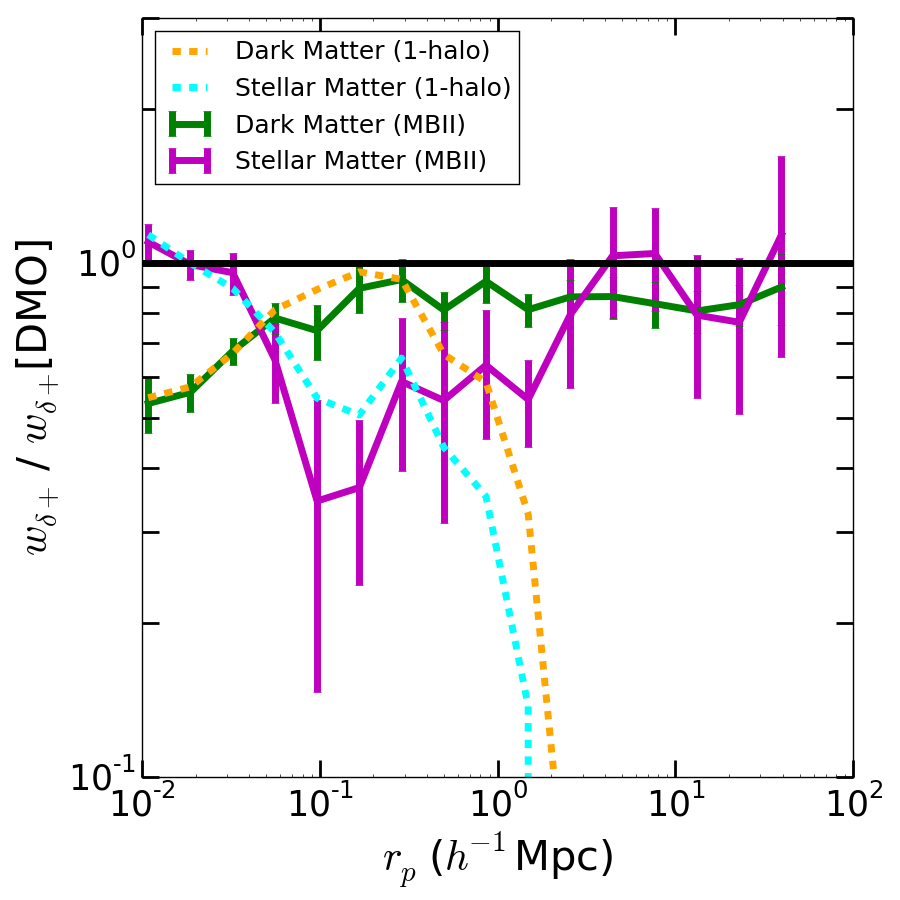}
\includegraphics[width=2.25in]{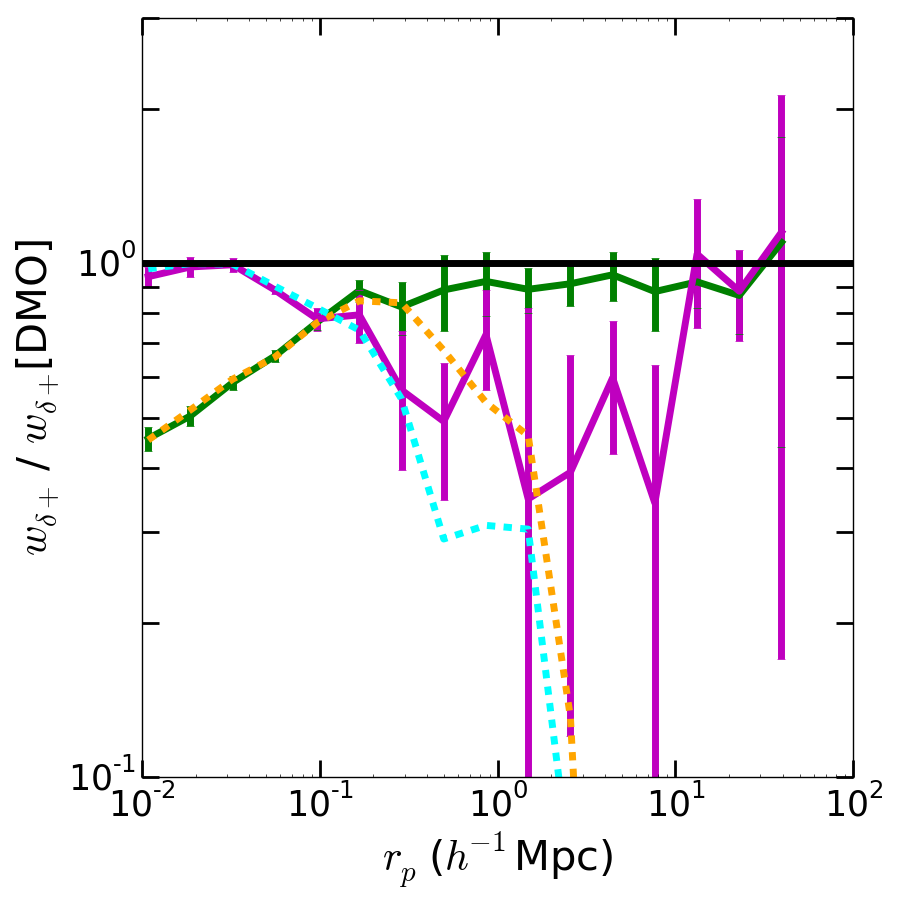}
\includegraphics[width=2.25in]{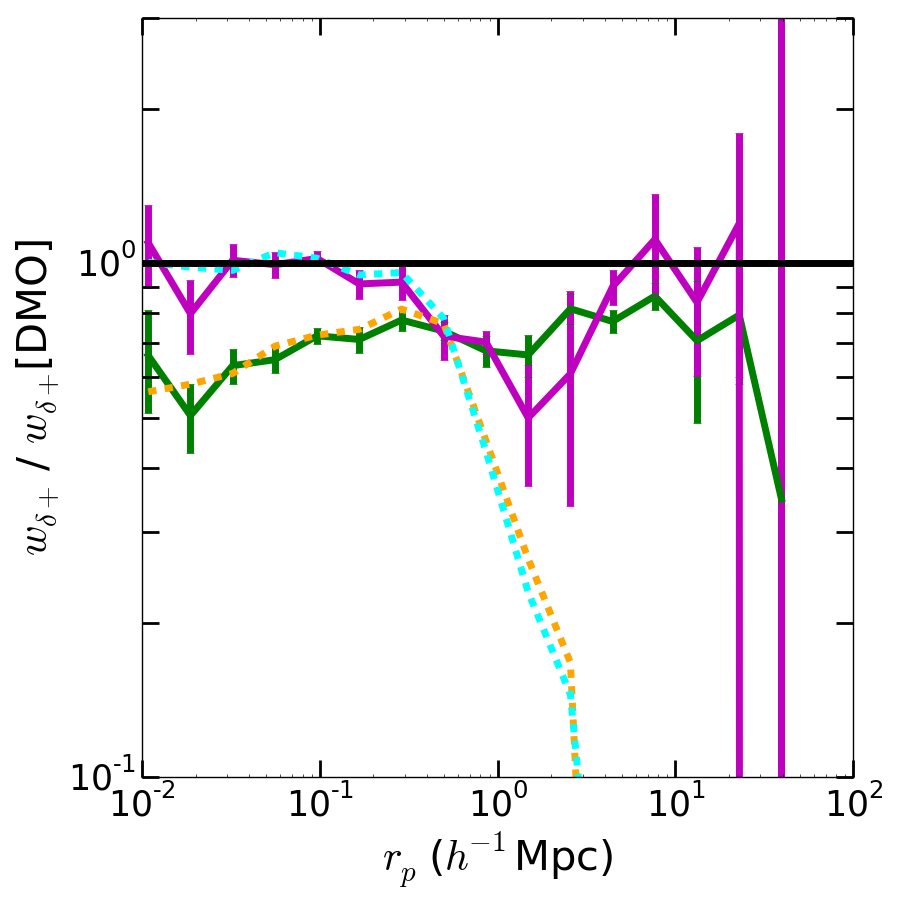}
\caption{\label{F:fig_wdeltap} Comparison of the projected density-shape correlation, $w_{\delta
    +}(r_{p})$, for the dark matter subhalo and stellar matter components in MBII with respect to
  that for dark matter subhalos in the DMO simulation,  computed separately for M1 (left), M2
  (midddle), and M3 (right).}
\end{center}
\end{figure*}

In this section, we compare the intrinsic alignments in MBII and DMO simulations by
analysing the two-point statistics, ellipticity-direction (ED) and the
projected density-shape ($w_{\delta +}$) correlation functions defined in Section~\ref{SS:pes}. In
Figure~\ref{F:fig_ed}, we compare the ED correlation for the shapes of the dark
matter subhalo and stellar matter component in MBII with that for 
the dark matter subhalos in the DMO simulation, mass
bins M1 ($10^{10.8-11.5}\hMsun$), M2 ($10^{11.5-13.0}\hMsun$), and M3 ($>10^{13.0}\hMsun$). The ED
correlation function for the dark matter subhalos in MBII is comparable to that
in the DMO simulation on large scales. However, on
small scales, we observe a tens of percent decrease in the ED correlation function for
the MBII simulation. This finding is possibly due to the dark matter subhalos in MBII being rounder
than in DMO, which implies that the dark
matter particles assume a distribution within the halos that is closer to spherical, reducing the ED
correlation function at smaller scales.

 The ED correlation computed using the shape of the stellar matter component in MBII is smaller than
 that of the dark matter component in DMO (by tens of percent, on all scales). This result is due to
 the misalignment of the stellar shape with the dark matter subhalo shapes that are determined by
 the density field. This ratio is relatively scale-independent in the higher mass bins M2 and M3,
 unlike the ratio of ED for the dark  matter subhalos in MBII vs.\ in DMO. On $0.5 - 5
   \hmpc$ scales, the fractional difference of this ratio is on average $\sim 72 \%$, $53 \%$ and $29\%$ in the mass bins M1, M2, and M3, respectively. 

In Fig.~\ref{F:fig_wdeltap}, we compare the projected shape-density
correlation function, $w_{\delta +}$, for the dark matter and stellar matter
shapes in MBII with that of the dark matter shape in DMO. Here, $w_{\delta +}$ for the dark matter
shape in DMO is higher than the other correlations at all values of $r_p$ and for all mass
bins. This finding is expected when comparing the DMO and MBII dark matter subhalos, since the
alignment of the dark matter shape with the density field is similar in both
simulations, but the dark matter subhalos in MBII are rounder, reducing $w_{\delta
  +}$ for MBII. If we consider the $w_{\delta +}$ computed using the shapes of the stellar
matter components, we observe that at small scales, it is close to the $w_{\delta
  +}$ of the dark matter shapes in the DMO simulation. This similarity derives from the compensation
of two competing effects: the stellar shapes are more misaligned with the density field, lowering
$w_{\delta +}$, but the stellar shapes have a larger ellipticity, raising $w_{\delta
  +}$. However, the correlation function computed using stellar component shapes is still $\sim 30-40 \%$ smaller than that using DMO
  subhalos for $0.5 <r_p< 5 \hmpc$. Due to the limited size of the simulation volume, the uncertainties are large beyond $\sim 5\hmpc$. At intermediate scales, we can also see the transition from
the $1$-halo term to the $2$-halo term for the stellar component, whereas for the dark matter component, it
is not clearly evident. To illustrate further, we also show the ratio for the $1$-halo terms corresponding to $w_{\delta +}$ of the shapes of dark matter and stellar component in MBII. 
In the lowest mass bin, which has a fairly equal mix of central and satellite subhalos, we observe a
change in the shape of the $1$-halo term at $\sim 1\hmpc$ for the stellar component.  
As the stellar component is more aligned with the
inner regions of the dark matter distribution in the subhalo, the $1$-halo
term drops more significantly in the intermediate scales in comparison to that
of the dark matter component.   

\begin{figure*}
\begin{center}
\includegraphics[width=2.25in]{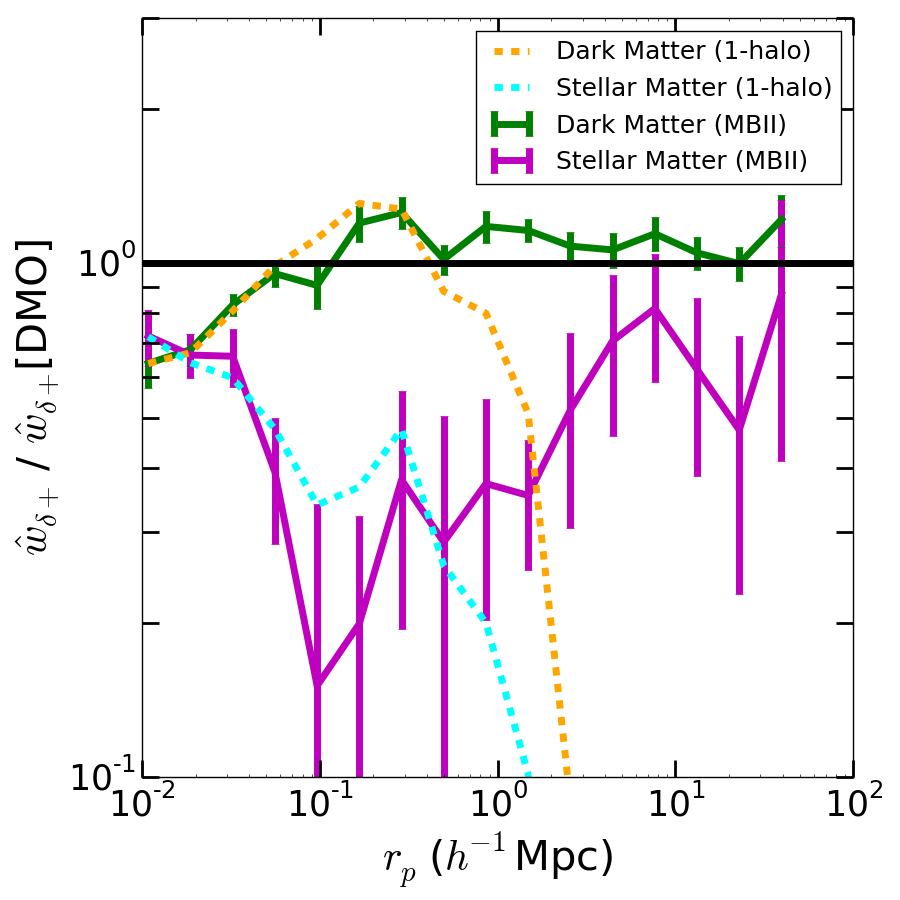}
\includegraphics[width=2.25in]{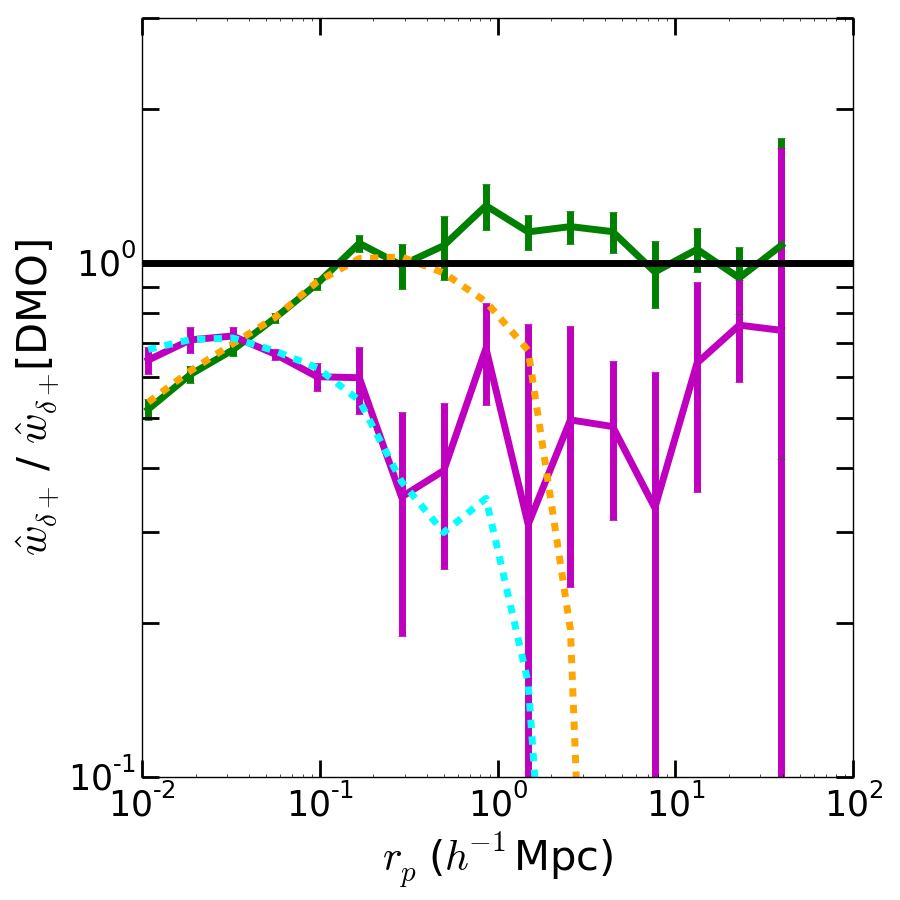}
\includegraphics[width=2.25in]{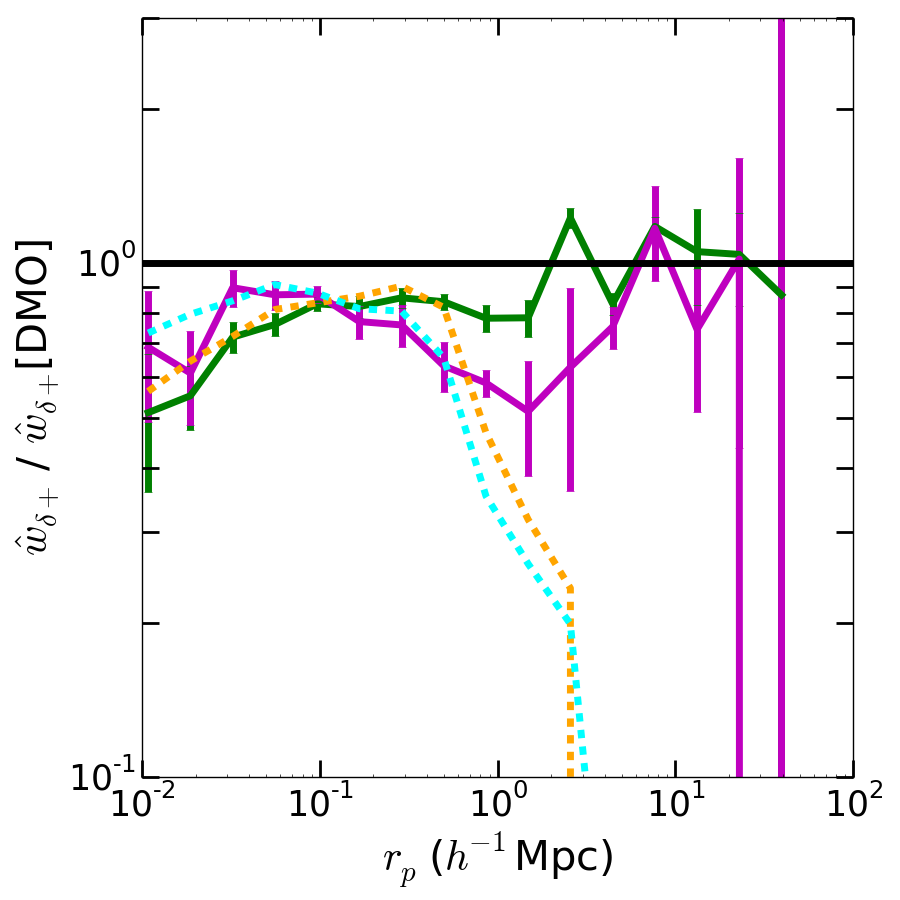}
\caption{\label{F:fig_wdeltapq0} Comparison of $\hat{w}_{\delta +}(r_{p})$ (projected shape-density correlation function without ellipticity weighting), for the dark matter subhalo and stellar matter components in MBII with respect to
  that for dark matter subhalos in the DMO simulation,  computed separately for M1 (left), M2
  (midddle), and M3 (right).}
\end{center}
\end{figure*}

For a more direct understanding of the effects of alignment versus
different shape distributions, we compute a new statistic, 
$\hat{w}_{\delta +}$, which is defined as the projected shape-density correlation
function without ellipticity weighting (i.e., setting $|e|=1$ for all objects in the shape sample).
The results are shown in Figure~\ref{F:fig_wdeltapq0}. For the stellar component in MBII, $\hat{w}_{\delta +}$ is
smaller than that of the dark matter component in DMO at all
scales. For $0.5 <r_p< 5 \hmpc$, the fractional differences in the correlation function are $\sim 60\%$, $49\%$ and $38\%$ in the mass bins M1, M2, and M3, respectively. Clearly, since the shapes have been normalized to the same value, this must be due only to
misalignments between galaxy shapes and the density field.  Similarly, the correlation
functions for the dark matter subhalo shapes agree on large scales, but that for MBII is smaller
than that for DMO on
small scales, due to the rounder shapes of the dark matter subhalos (just as was seen for the ED
correlation). 

\section{Conclusions}\label{S:conclusions}

In this paper, we carried out a comparison of halo and subhalo populations in the MassiveBlack-II
(MBII) simulation, and a corresponding dark matter-only simulation (DMO) with the same cosmology,
resolution, volume, and initial conditions.  First, considering basic
halo properties, we compared the halo mass function of the MBII simulation with that of DMO. The
mass function is suppressed in the hydrodynamic run by $\sim 10 \%$ at $10^{12}\hMsun$, increasing
to $20 \%$ at $10^{9}\hMsun$. This agrees qualitatively with the findings from the OWLS simulation
\citep{2014MNRAS.442.2641V}. The two-point correlation function for the dark matter particles in the
two simulations is similar on large scales. On small scales, the two-point correlations are larger
for the hydrodynamic run, corresponding to an increased dark matter power spectrum at large $k$
values. 

Identifying matching subhalos in the two simulations enables us to directly compare the distribution
of axis ratios and shape alignments to understand the effects of baryonic physics.  The fraction of
matched subhalos decreases as we go to lower masses, but is above 90 per cent for the mass range
used in this work.  These shapes are also used to calculate and compare the intrinsic alignment
two-point statistics.

We measured the shapes of dark matter  subhalos in MBII and DMO simulations as a function of radius,
by only using particles within a certain distance from the center based on a fixed fraction of the
subhalo $R_{200}$. We analyzed the distributions of axis ratios and alignments of shapes measured
within  $0.2 R_{200}$, $0.6 R_{200}$, and $1.0 R_{200}$. In both simulations, we found that the axis
ratios increase with the distance from center, which agrees with previous results in $N$-body
simulations \citep{2012JCAP...05..030S}. 
We also found that the dark matter subhalo axis ratios in the MBII simulation are higher (rounder)
than those in the DMO simulation at all mass ranges due to the effects of baryonic physics, in
agreement with previous findings
\citep[e.g.,][]{{2004ApJ...611L..73K},{2010MNRAS.407..435A},{2011MNRAS.415.2607D}}. The galaxy
stellar components in MBII have smaller axis ratios than the dark matter subhalos in both
simulations, with the fractional difference being larger for the minor-to-major axis ratio,
$s$. 

The degree of alignment of the stellar component in MBII with the dark matter component in DMO is
larger in subhalos of high mass and decreases at lower masses. This trend is qualitatively similar
to that of the alignment of the stellar component with the dark matter component in MBII. In
subhalos of higher mass, the shapes of the dark matter component in MBII and DMO simulation are well
aligned with each other. This alignment is stronger than the alignment of the dark matter shape in
MBII with that of stellar component in the same simulation. Comparing the misalignment of the
stellar component with shapes of dark matter component measured within different radial distances in
MBII, we find that the misalignment angles increase as we go to larger radii within the dark matter
subhalo. However, we do not notice a significant change in the misalignment of the stellar component
with the shape of dark matter component in DMO measured for different radii. Based on our results
from axis ratios and misalignment angles, we concluded that when mapping galaxy alignments from
hydrodynamic simulations onto subhalos in dark matter-only simulations, it is not useful to measure the shape of dark matter component at smaller radii in order to trace the shape and orientation of the stellar component. Hence, for the comparison of two-point statistics, we only consider the shapes of dark matter subhalos obtained using all the particles in the subhalo. 

Using the shapes of matched subhalos, we compared the intrinsic alignments two-point statistics
(ellipticity-direction correlation, or ED, and the projected-shape density correlation, $w_{\delta
  +}$)  of the dark matter and stellar matter in MBII with that of the dark matter component in DMO
simulation. The ED correlations of the dark matter component in MBII and DMO agree on large
scales. On small scales, the ED correlation function decreases in MBII due to a change in the dark
matter profile caused by baryonic physics. For the stellar component, the ED correlation is smaller
on all scales due to the misalignment of the stellar component with the dark matter component. This corresponds to a fractional difference ranging from $\sim 30 - 70 \%$ for scales around $0.5-5 \hmpc$, and the   
decrease is larger in subhalos of lower mass. The $w_{\delta + }$ correlation function  for the
shapes of dark matter component in MBII are smaller when compared to that of the DMO simulation due
to the smaller values of ellipticities, since the shapes are rounder in MBII. For the stellar
component, we find that the $w_{\delta +}$ is comparable on large scales and small scales with the
$w_{\delta +}$ of the dark matter component in DMO. However, for scales around $0.5-5 \hmpc$, $w_{\delta +}$ is still smaller for the stellar component by $\sim 30 - 40 \%$. At intermediate scales, we find a transition
from the $1$-halo to $2$-halo term that causes a decrease in $w_{\delta +}$ computed using stellar
shapes, but this feature is not clearly evident for the $w_{\delta +}$ of the dark matter shape in
MBII.

Our results in this paper suggest that the scatter in the distribution of axis ratios of the dark
matter subhalo shapes in the DMO simulation is large compared to that of the stellar component in
MBII, with significant misalignment in the orientation of shapes. However, the alignments between
the galaxies in MBII and the corresponding matched subhalos in the DMO simulation are still
significant compared to a uniform random distribution.  In future work, we will  use these
measurements to map the intrinsic alignments of the stellar matter component in hydrodynamic
simulations onto dark-matter-only simulations.     This is an important step in producing $N$-body
based mock catalogs that have realistically-complicated intrinsic alignments for tests of weak
lensing analysis methods in future surveys.

\section*{Acknowledgments}

AT and RM acknowledge the support of NASA ROSES 12-EUCLID12-0004. We thank Rachel Bean,
Jonathan Blazek, Nick Gnedin, Katrin Heitmann, Michael Schneider and other member of the LSST-DESC collaboration for providing helpful
feedback on this work.  TDM has been funded by the National Science Foundation (NSF)
PetaApps, OCI-0749212 and by NSF AST-1009781 and ACI-1036211. AK was supported in part by JPL, run under a contract by Caltech for NASA. AK was also supported in part by NASA ROSES 13-ATP13-0019. 
This research used resources of the National Energy Research Scientific Computing Center (NERSC), a DOE
Office of Science User Facility supported by the Office of Science of the U.S. Department of Energy
under Contract No.\ DE-AC02-05CH11231. 

\bibliographystyle{mn2e2} \bibliography{draft3.bib}


\end{document}